\documentclass[a4paper,11pt,reqno]{article}

\usepackage{graphicx,epstopdf}
\usepackage{amsmath, amsfonts, amssymb}
\usepackage[english]{babel}
\usepackage[breaklinks=true]{hyperref}
\usepackage{lpic}

\advance\textwidth by 1.4in\advance\oddsidemargin by - 0.7in
\advance\textheight by 1.1in\headheight 0pt\topskip 0pt\headsep 0pt\topmargin 0pt

\hypersetup{
			colorlinks=true,
			urlcolor=blue,
			citecolor=magenta,
			linkcolor=blue,
     }
\numberwithin{equation}{section}
\let\oldsqrt\sqrt
\def\sqrt{\mathpalette\DHLhksqrt}
\def\DHLhksqrt#1#2{%
\setbox0=\hbox{$#1\oldsqrt{#2\,}$}\dimen0=\ht0
\advance\dimen0-0.2\ht0
\setbox2=\hbox{\vrule height\ht0 depth -\dimen0}%
{\box0\lower0.4pt\box2}}

\begin{document}

\begin{center}

\vspace{.7cm}
{\LARGE\bf Viscous Asymptotically Flat \\[1mm] Reissner-Nordstr\"om Black Branes} \\

\vspace{1.2cm}

{{\textbf{Jakob Gath} and \textbf{Andreas Vigand Pedersen}}\\
\vspace{1cm}
 {\small\slshape Niels Bohr Institute,\\
 University of Copenhagen,\\
 Blegdamsvej 17, DK-2100 Copenhagen \O,
 Denmark}}\\
 
\vspace{0.6cm}

{\small  \texttt{gath@nbi.dk, vigand@nbi.dk}}
\vspace{1.5cm}

\thispagestyle{empty}

{\bf Abstract} \end{center} \vspace{0mm} {We study electrically charged asymptotically flat black brane solutions whose world-volume fields are slowly varying with the coordinates. Using familiar techniques, we compute the transport coefficients of the fluid dynamic derivative expansion to first order. We show how the shear and bulk viscosities are modified in the presence of electric charge and we compute the charge diffusion constant which is not present for the neutral black $p$-brane. 
We compute the first order dispersion relations of the effective fluid. For small values of the charge the speed of sound is found to be imaginary and the brane is thus Gregory-Laflamme unstable 
as expected. For sufficiently large values of the charge, the sound mode becomes stable, however, in this regime the hydrodynamic mode associated with charge diffusion is found to be unstable. The electrically charged brane is thus found to be (classically) unstable for all values of the charge density in agreement with general thermodynamic arguments.
Finally, we show that the shear viscosity to entropy bound is saturated, as expected, while the proposed bounds for the bulk viscosity to entropy can be violated in certain regimes of the charge of the brane.}

\newpage
\clearpage
\pagenumbering{arabic}

\begingroup
\hypersetup{linkcolor=black}
\tableofcontents
\endgroup
\newpage

\section{Introduction}

Black branes possess hydrodynamic properties in addition to their thermodynamic properties. Indeed, fluid dynamics is the natural local generalization of the global thermodynamics of black branes. This is well-known and established in the hydrodynamic limit of the AdS/CFT correspondence \cite{Bhattacharyya:2008jc}, but has also been shown to hold true for Ricci flat black branes \cite{Emparan:2009cs, Emparan:2009at, Camps:2010br, Camps:2012hw}. Here the effective dynamics of the black brane (collectively known as 'the blackfold approach') is captured by the quasi-local stress tensor of Brown and York \cite{Brown:1992br}. In general, the stress tensor is given
by a fluid-elastic derivative expansion in the collective parameters describing the brane. To leading order, the stress tensor is that of a boosted perfect fluid which accounts for the thermodynamics of the brane. Fluctuations in the directions transverse to the brane (i.e. bending) give rise to elastic contributions \cite{Emparan:2007wm, Armas:2011uf, Armas:2012ac, Armas:2012jg} while 
fluctuations in the world-volume fields longitudinal to the brane, relevant to this paper, are captured by fluid dynamic dissipative corrections to the effective stress tensor \cite{Camps:2010br}. 

In this paper we examine how the presence of electric Maxwell charge modifies the fluid transport coefficients of the neutral black brane originally considered in \cite{Camps:2010br}. Moreover, we compute the transport coefficient associated to charge diffusion in the charged black brane to first order in the fluid derivative expansion. Although many supergravity solutions carrying lower dimensional charge on their world-volume are known \cite{Costa:1996zd, Breckenridge:1996tt}, only recently a new family of black branes of Einstein gravity coupled to a dilaton  and a single $q+1$ form gauge field was found \cite{Caldarelli:2010xz}. This family of solutions has both the dilaton coupling and the (integer) dimension $q$ as free parameters. The Maxwell charged black brane solution can therefore straightforwardly be obtained from this family of solutions by specializing to $q=0$ and turning off the dilaton coupling. This especially means that we can study hydrodynamic fluctuations of pure Einstein gravity coupled to a single gauge field without a dilaton - something which is not possible for generic supergravity backgrounds. In this way, our computation provides the simplest generalization of the neutral case. In detail, we consider long wavelength fluctuations around the black brane solution of Einstein-Maxwell gravity (which we shall from now on dub the Reissner-Nordstr\"{o}m black brane) following the method of \cite{Bhattacharyya:2008jc, Camps:2010br}. We solve the full set of coupled Einstein-Maxwell equations to first order in the derivative expansion and compute the effective stress tensor and current. This provides us with the charged generalizations of respectively the shear and bulk viscosity along with the charge diffusion constant which is not present in the neutral case. 

Having computed the shear and bulk viscosities we find that the bound $\eta/s\geq 1/4\pi$ is saturated. This agrees with the expectation that this should hold for any two-derivative gravity theory \cite{Policastro:2001yc, Buchel:2003tz}. However, we find that the 
bulk viscosity to entropy bound proposed in \cite{Buchel:2007mf} can be violated in certain regimes. Perhaps this is not too surprising since the derivation of this bound relies heavily on
holographic considerations. However, it is worth noting that the bound is saturated for the neutral black brane \cite{Camps:2010br}. Finally we mention that the modified bound proposed in \cite{Gouteraux:2011qh, Smolic:2013gx} is found to be violated. 

By computing the speed of sound in the effective fluid of the neutral black brane ref. \cite{Emparan:2009at} was able to identify the unstable sound mode of the effective fluid with the Gregory-Laflamme (GL) instability \cite{Gregory:1994bj, Harmark:2007md}. This (very simple) computation can already be
carried out at the perfect fluid level. The results of \cite{Camps:2010br} allowed further refinement of this result and showed remarkable agreement with numerical data. Performing a similar computation for the Reissner-Nordstr\"{o}m black brane, we find that for a given temperature, the speed of sound becomes real above a certain threshold value of the charge density. This seems to imply that the Reissner-Nordstr\"{o}m black brane is not GL unstable for sufficiently large values of the charge. However, upon closer inspection of (next-to-leading order) dispersion relations, we find that above the threshold value of the charge, the hydrodynamic mode associated with charge diffusion becomes unstable. The Reissner-Nordstr\"{o}m black brane is therefore leading order and next to leading order GL unstable below and above the charge threshold, respectively. This complementary behavior of the instability is expected from the general thermodynamics. Indeed, by computing the specific heat $C$ and the isothermal electric permittivity $c$ we find that the conditions $C>0$ and $c>0$ put complementary conditions on the charge density of the brane (recall that $C<0$ for the neutral brane). Since thermodynamic stability requires that $C>0$ and $c>0$, the Reissner-Nordstr\"{o}m black brane is never thermodynamically stable (see e.g. \cite{Harmark:2005jk}).

In many ways, studying intrinsic  fluctuations of branes in the blackfold formalism is similar in spirit to the well-known fluid/gravity correspondence of AdS/CFT \cite{Bhattacharyya:2008jc}. We also mention that the computation in fluid/gravity analogous to ours (fluctuations of the AdS Reissner-Nordstr\"{o}m brane of co-dimension 1 with a Chern-Simons term) was carried out in the papers \cite{Maeda:2006by, Son:2006em, Banerjee:2008th, Erdmenger:2008rm}. However, we emphasize that our computation deals with asymptotically flat branes of general co-dimension (note that extensions of the blackfold formalism to AdS backgrounds have been made in certain regimes - see \cite{Caldarelli:2008pz, Camps:2008hb, Armas:2010hz, Armas:2012bk}) and that the effective fluid stress tensor has no direct interpretation as a "dual" fluid of a QFT. Also note that the blackfold fluid stress tensor is not that of a conformal fluid. 
Recently a connection between the fluid/gravity correspondence and the blackfold formulation was established. This was done by constructing a map from asymptotically AdS solutions compactified on a torus to a corresponding Ricci-flat solution obtained by replacing the torus by a sphere \cite{Caldarelli:2012hy}. This was used to take the general second order results of fluid/gravity \cite{Bhattacharyya:2008mz} and map them to the second order blackfold stress tensor. This provides even further improvement of the dispersion relation of the GL instability.

The outline of the paper is as follows: In the subsequent section \ref{sec:start} we will setup the leading order solution and review its thermodynamics. In section \ref{sec:perb} we discuss the perturbation procedure and explain how the boundary conditions are handled. In section \ref{sec:firstorder} the first order equations are solved and in section \ref{sec:stresstensor} the effective stress tensor and current are provided. In section \ref{sec:stability} the transport coefficients are used to analyze the dispersion relations and the GL instability. Finally, we will discuss the results in section \ref{sec:discuss} and address some interesting future developments.

\paragraph{Notation:}

We use $\mu,\nu$ to label the $D=p+n+3$ spacetime directions. Moreover, we denote the $p+1$ world-volume directions of the brane in Schwarzschild coordinates by $x^a = (t,x^i)$ and in Eddington-Finkelstein coordinates by $\sigma^a = (v, \sigma^i)$ with $a=0 \ldots p$ and $i=1, \ldots, p$. The co-dimension of the brane is $n+2$. For simplicity of the presentation we restrict ourselves to the cases where $n>1$ due to a slightly different behavior at infinity for the $n=1$ solution. However, treating the special case of $n=1$ should be straightforward using similar considerations as for the neutral case. 

\section{Reissner-Nordstr\"{o}m branes and effective zeroth order fluid} \label{sec:start}

In this section we review the generalized Gibbons-Maeda solution for $q=0$ which was found in \cite{Caldarelli:2010xz}. The generalized Gibbons-Maeda solution describes a black $p$-brane with horizon topology $S^{n+1}\times\mathbb{R}^p$ which has electric $q$-charge diluted on its world-volume. The solution was  obtained from the Gibbons-Maeda solution \cite{Gibbons:1987ps} through an elaborate double uplifting procedure. The general solution is given in terms of a metric, a dilaton and a $(q+1)$-form gauge field under which (the $q$-charge diluted on the) black $p$-brane is charged. A particularly nice property of the generalized Gibbons-Maeda solution is that the dilaton coupling $a$ can be treated as a free parameter. This especially means that we are free to set $a=0$. This is of course not possible for the well-known supergravity solutions such as the D$0$-D$p$ system \cite{Costa:1996zd, Breckenridge:1996tt}. Moreover, in this paper we
restrict ourselves to the $q=0$ case (Maxwell charge).

\subsection{Reissner-Nordstr\"{o}m black branes} 
As explained above, we consider branes of Einstein-Maxwell theory. The action is
\begin{equation} \label{eqn:theory}
S = \frac{1}{16\pi G} \int d^Dx\sqrt{-g}\left[R-\frac{1}{4}F_{\mu \nu}F^{\mu\nu}\right] \,,
\end{equation}
where $F_{\mu \nu}$ is the field strength of the Maxwell gauge field $A_\mu$, $F=\text{d}A$. We now present the Reissner-Nordstr\"{o}m black brane solution. The solution is characterized by $p$ flat spatial directions $x^i$, a time direction $t$, and finally a radial direction $r$ along with the usual transverse sphere $S^{n+1}$. The total spacetime dimension $D$ is related to
$p$ and $n$ by $D=p+n+3$. The metric is given by 
\begin{equation} \label{eqn:RNmetric}
\text{d}s^2=-h^{-2}f \thinspace \text{d}t^2+h^B\left(f^{-1}\text{d}r^2+r^2\text{d}\Omega_{(n+1)}^2+ \sum_{i=1}^p\left(\text{d}x^i\right)^2 \right) \,.
\end{equation}
The two harmonic functions $f\equiv f(r)$ and $h\equiv h(r)$ are given by\footnote{In the blackfold literature $\gamma_0 \equiv \sinh^2\alpha$.}
\begin{equation}
f(r)=1-\left(\frac{r_0}{r}\right)^n, \quad h(r)=1+\left(\frac{r_0}{r}\right)^n\gamma_0 \,.
\end{equation}
The two parameters $r_0$ and $\gamma_0$ are related to the thermal and electrostatic energy of the solution (see below). The parameter $B\equiv B(p,n)$ is given by\footnote{The full generalized Gibbons-Maeda solution
has an additional parameter $A$. However, for the non-dilatonic Reissner-Nordstr\"{o}m solution one has $A=2$.}
\begin{equation}
B=\frac{2}{n+p} \,.
\end{equation}
Finally the gauge field $A$ is given by  
\begin{equation}
A = -\frac{\sqrt{N}}{h}\left(\frac{r_0}{r}\right)^n\sqrt{\gamma_0(\gamma_0+1)} \thinspace \text{d}t \,,
\end{equation}
where we have defined $N\equiv B+2$.

It is straightforward to apply a uniform boost $u^a$ to the solution \eqref{eqn:RNmetric} in the flat directions: The metric of the non-dilatonic boosted Reissner-Nordstr\"{o}m black brane is given by
\begin{equation} \label{eqn:leadGMmetric}
\text{d}s^2=h^B\left(-h^{-N} f \thinspace u_a u_b \thinspace \text{d}x^a\text{d}x^b+f^{-1}\text{d}r^2+r^2\text{d}\Omega_{(n+1)}^2+ \Delta_{ab} \thinspace \text{d}x^a\text{d}x^b \right) \,.
\end{equation}
where $\Delta^{a}_{\ b} \equiv \delta^{a}_{\ b}+u^a u_b$ is the usual orthogonal projector defined by the boost $u^a$. The gauge field is given by
\begin{equation}
A = \frac{\sqrt{N}}{h}\left(\frac{r_0}{r}\right)^n\sqrt{\gamma_0(\gamma_0+1)} \thinspace u_a \text{d}x^a \,.
\end{equation}
In the next section we review the thermodynamics and effective blackfold description of this solution.

\subsection{Thermodynamics and effective blackfold fluid} \label{sec:termosec}

The blackfold theory of $p$-branes supporting $q=0$ charge was developed in \cite{Caldarelli:2010xz} and further examined in \cite{Emparan:2011hg} where also $p=q$ was considered in detail (in a supergravity setting). For a uniform boost $u^a$ of the brane, it is instructive to write the effective blackfold stress tensor 
in the form 
\begin{equation}\label{genstresstensor}
  T^{ab}_{(0)}=\mathcal T s \left(u^a u^b-\frac{1}{n}\gamma^{ab} \right)+\Phi \mathcal{Q} \thinspace u^a u^b \,,
\end{equation}
where $\gamma_{ab}$ is the induced metric on the blackfold. For our purposes (flat extrinsic geometry), we have $\gamma_{ab}=\eta_{ab}$. Moreover $\mathcal T$ is the local temperature, $s$ is the entropy density, $\mathcal Q$ is the charge density and finally $\Phi$ is the electric potential conjugate to $\mathcal Q$. The various quantities are parameterized in terms of a charge parameter $\gamma_0$ and the horizon thickness $r_0$:
\begin{align}\label{parar0gamma0}
\begin{split}
\mathcal T& =\frac{n}{4\pi r_0 \sqrt{(1+\gamma_0)^N}}, \quad
s =\frac{\Omega_{(n+1)}}{4G}r_0^{n+1}\sqrt{(1+\gamma_0)^N}, \\
\mathcal Q& = \frac{\Omega_{(n+1)}}{16\pi G} n \sqrt{N} r_0^n \sqrt{\gamma_0 (1+\gamma_0)}, \quad
\Phi =\sqrt{\frac{N\gamma_0}{1+\gamma_0}} \,.
\end{split}
\end{align}
The stress tensor in the form \eqref{genstresstensor} immediately 
allows us to identify the thermal and the electrostatic parts. Since $r_0^n\sim \mathcal{T}s$, $r_0$ gives us
a measure of the thermal energy (density) of the given solution. In a similar manner $\gamma_0$ is identified with the thermodynamic ratio,
\begin{align}
\gamma_0 = \frac{1}{N} \frac{\Phi \mathcal Q}{ \mathcal T s} \,, 
\end{align}
and $\gamma_0$ therefore measures the electrostatic energy relative to the thermal energy of the black brane. 

Of course, it is straightforward to cast the stress tensor into 
standard form (using the Gibbs-Duhem relation $w\equiv \epsilon+P=\mathcal T s+\Phi \mathcal Q$)
\begin{equation}
T^{ab}_{(0)}=w u^a u^b+ P\eta^{ab}=\epsilon u^a u^b+P\Delta^{ab} \,,
\end{equation}
where
\begin{equation} \label{eqn:energydensity}
\epsilon=\frac{\Omega_{(n+1)}}{16\pi G} r_0^n \big(n+1 + nN\gamma_0 \big), \quad P=-\frac{\Omega_{(n+1)}}{16\pi G} r_0^n, \quad w=\frac{n\Omega_{(n+1)}}{16\pi G} r_0^n \big(1 + N\gamma_0 \big) \,.
\end{equation}
Finally the $q=0$ current supported by the $p$-brane is given by 
\begin{equation}
J^a_{(0)}=\mathcal Q \thinspace u^a \,.
\end{equation}
To leading order, the intrinsic blackfold equations take the form of the world-volume conservation equations $\nabla_a T^{ab}_{(0)}=0$ and $\nabla_a J^a_{(0)}=0$. For flat extrinsic geometry $\gamma_{ab}=\eta_{ab}$, 
they evaluate to the equations 
\begin{equation} \label{eqn:ConsEqns}
 \dot{\epsilon}=-w \vartheta, \quad \dot{u}^a= -w^{-1}\Delta^{ab}\partial_b P, \quad \dot{\mathcal{Q}}=-\mathcal{Q} \vartheta \,,
\end{equation}
where $\vartheta\equiv \partial_a u^a$ is the expansion of $u^a$ and a dot denotes the directional derivative along $u^a$. The (first order) equations will be important in the
perturbative analysis. As expected, they will show up as constraint equations when solving the Einstein-Maxwell system perturbatively.

\section{The perturbative expansion} \label{sec:perb}

As explained in the introduction, the aim of this paper is to solve the Einstein-Maxwell system in a derivative expansion around the solution given in section \ref{sec:start}. In this section we define the appropriate 
coordinates to handle this problem and explain how the perturbations are classified according to their transformation properties under $\text{SO}(p)$.

\subsection{Setting up the perturbation}

Before perturbing the brane, we first need to cast the metric \eqref{eqn:leadGMmetric} into Eddington-Finkelstein-like (EF) form. The reason is two-fold. First, it is essential for the computation that we can ensure regularity at the horizon and since the Schwarzschild description breaks down at the horizon, it is clearly more useful to use EF coordinates. Secondly, since a gravitational disturbance moves along null-lines, in order to control the perturbation, we want the lines of 
constant world-volume coordinates to be radial null-curves i.e. $g_{rr}=0$. This is exactly the defining property of EF coordinates. For a general boost $u^a$, we define the EF coordinates $\sigma^a$ by 
\begin{equation} \label{eqn:EFcoord}
\sigma^a=x^a+u^ar_\star, \quad r_\star(r)=r+\int_{r}^{\infty}\left(\frac{f-h^{N/2}}{f}\right)dr \,.
\end{equation}
Here $r_\star$ is chosen such that $r_\star'=h^{N/2}/f$ and $r_\star \to r$ for large $r$. The first condition ensures that $g_{rr}=0$ while the latter is chosen such that the EF coordinates reduce to ordinary
radial Schwarzschild light cone coordinates for large $r$. Notice that it is possible to write down a closed form expression for $r_\star$ in terms of the hypergeometric Appell function $F_1$  
\begin{equation}
r_\star(r)=r F_1 \left(-\frac{1}{n};-\frac{N}{2},1;1-\frac{1}{n};1-h,1-f\right)\approx r\left(1-\frac{1}{n-1} \frac{r_0^n}{r^n} \left(1+\frac{N\gamma_0}{2}\right)\right) \,,
\end{equation}
where the last equality applies for large $r$ and is valid up to $\mathcal{O}\left(\frac{1}{r^{2n-1}}\right)$. It is nice to note that the hypergeometric Appell function $F_1$ reduces to the ordinary hypergeometric function $_2 F_1$ in the neutral limit $\gamma_0\to 0$. Indeed
\begin{equation}
\lim_{\gamma_0\to 0} r_\star(r)=r_\star(r)\Big|_{\gamma_0=0}\equiv r+\int_{r}^\infty\left(\frac{f-1}{f}\right)dr=r \thinspace _2F_1\left(1;-\frac{1}{n};1-\frac{1}{n}; 1-f \right) \,,
\end{equation}
which is the $r_\star$ used in \cite{Camps:2010br}. With this definition of $r_\star$ we will limit our analysis to the case for which $n \geq 2$. In EF coordinates, the metric \eqref{eqn:leadGMmetric} takes the form 
\begin{equation}
\text{d}s^2_{(0)}=h^B\left(-h^{-N}f \thinspace u_a u_b \thinspace \text{d}\sigma^a\text{d}\sigma^b-2h^{-N/2}\thinspace u_a \thinspace \text{d}\sigma^a \text{d}r+\Delta_{ab} \thinspace \text{d}\sigma^a\text{d}\sigma^b+r^2\text{d}\Omega_{(n+1)}^2\right) \,.
\end{equation}
Here the subscript indicates that the metric solves the Einstein-Maxwell equations to zeroth order in the derivatives. Notice that in these coordinates the gauge field will acquire a non-zero $A_r$ component . However, we shall work in a gauge where this component is zero. We therefore take 
\begin{equation} A^{(0)}=\frac{\sqrt{N}}{h}\left(\frac{r_0}{r}\right)^n\sqrt{\gamma_0(\gamma_0+1)} \thinspace u_a \text{d}\sigma^a \,, \quad \text{and in particular} \quad A_r^{(0)}=0 \,, 
\end{equation}
Having determined the EF form of the metric and gauge field, we are now ready to set up the perturbative expansion.\vspace{5mm}
 
Following the lines of \cite{Camps:2010br}, we promote the parameters $u^a, r_0$ and $\gamma_0$ to \emph{slowly} varying world-volume fields: 
\begin{equation}
u^a\to u^a(\sigma^a), \quad r_0\to r_0(\sigma^a), \quad \gamma_0\to \gamma_0(\sigma^a) \,.
\end{equation}
By slowly varying we mean that the derivatives of the world-volume fields are sufficiently small. In order to quantify this, we introduce a set of re-scaled coordinates $\sigma^a_\varepsilon=\varepsilon \sigma^a$, $\varepsilon \ll 1$, and consider the w.v. fields to be functions of $\sigma^a_\varepsilon$. In this way each derivative will produce a factor of $\varepsilon$. Moreover, two derivatives will be suppressed by a factor of $\varepsilon$ compared to one derivative and so on. Effectively what we are doing is to consider arbitrary varying world-volume fields (no restrictions on the size of derivatives) and ``stretching'' them by a factor of $1/\varepsilon\gg 1$. In this way we will only consider slowly varying fields and the derivative expansion is controlled by the parameter $\varepsilon$.\footnote{In the end of the computation, we of course set $\varepsilon=1$ and keep in mind that the expressions only hold as a derivative expansion i.e. for sufficiently slowly varying configurations.} The fields can now be expanded around a given point $\mathcal P$
\begin{equation}
\begin{split}
  u^a(\sigma) = u^a\big|_\mathcal{P}+\varepsilon \sigma^b \partial_b u^a&|_\mathcal{P} + \mathcal{O}(\varepsilon^2) \,, \quad  %
  r_0(\sigma) = r_0\big|_\mathcal{P}+\varepsilon \sigma^a\partial_a    
r_0|_\mathcal{P}+\mathcal{O}(\varepsilon^2) \,, \\ %
  \gamma_0(\sigma) &= \gamma_0\big|_\mathcal{P}+\varepsilon \sigma^a\partial_a \gamma_0|_\mathcal{P}+\mathcal{O}(\varepsilon^2) \,.
\end{split}
\end{equation}
We now seek derivative corrections to the metric and gauge field denoted by respectively $\text{d}s_{(1)}^2$ and $A_{(1)}$, so that 
\begin{equation}
\text{d}s^2=\text{d}s^2_{(0)}+\varepsilon \text{d}s^2_{(1)}+\mathcal{O}(\varepsilon^2) \quad \text{and} \quad A=A_{(0)}+\varepsilon A_{(1)}+\mathcal{O}(\varepsilon^2) \,,  
\end{equation}
solves the equations of motion to order $\varepsilon$. By a suitable choice of coordinates, we can take the point $\mathcal P$ to lie at the origin $\sigma^a=(0,\mathbf{0})$. Moreover, we can choose coordinates so that $u^v\big|_{(0,\mathbf{0})}=1$, $u^i\big|_{(0,\mathbf{0})}=0$, $i=1,...,p$ (the rest frame of the boost in the origin)\footnote{In these coordinates $u^v=1+\mathcal{O}(\varepsilon^2)$.}. In these particular coordinates, the
$0^{\text{th}}$ order metric $\text{d}s^2_{(0)}$ takes the form 
\begin{equation}
\begin{split}
	\text{d}s^2_{(0)} &= h^B \left[ - 2  h^{-\frac{N}{2}} \text{d}v\text{d}r - \left( \frac{f}{h^{N}} \right) \text{d}v^2 + \sum_{i=1}^{p} (\text{d}\sigma^i)^2 + r^2 \text{d}\Omega_{(n+1)}^2 \right]  \\
	& + \varepsilon h^B  \Bigg[ \frac{1}{h^N} \frac{r_0^n}{r^n} \left( \frac{n}{r_0} \left( 1 + \frac{2f}{h} \gamma_0 \right)  \sigma^a \partial_a r_0 + \frac{2f}{h} \sigma^a \partial_a \gamma_0 \right) \text{d}v^2   \\
	& + \frac{B}{h}\frac{r_0^n}{r^n} \left(  \frac{n  \gamma_0 }{r_0} \sigma^a \partial_a r_0 + \sigma^a \partial_a \gamma_0 \right) \left( \sum_{i=1}^{p} (\text{d}\sigma^i)^2 + r^2 \text{d}\Omega_{(n+1)}^2 \right)   \\
	& + 2 \left( \frac{f}{h^N} - 1 \right) \sigma^a \partial_a u_i \thinspace \text{d}v\text{d}\sigma^i
	- \frac{2}{h^{N/2}} \sigma^a \partial_a u_i \thinspace \text{d}\sigma^i\text{d}r   \\
	& + \frac{B-2}{h^{N/2+1}}\frac{r_0^n}{r^n} \left(  \frac{n \gamma_0 }{r_0}  \sigma^a \partial_a r_0 + \sigma^a \partial_a \gamma_0 \right) \text{d}v\text{d}r  
	%
	%
	 \Bigg]  \,,
\end{split}
\end{equation}
where we have denoted $r_0|_{(0,\mathbf{0})} \equiv r_0$ and $\gamma_0|_{(0,\mathbf{0})} \equiv \gamma_0$. Clearly the system has a large amount of gauge freedom. Following the discussion of the definition of $r_\star$, we want the $r$ coordinate to maintain its geometrical interpretation. We therefore choose
\begin{equation}
g^{(1)}_{rr}=0  \,,
\end{equation}
and we moreover take 
\begin{equation} \label{eqn:gaugechoice2}
 g^{(1)}_{\Omega\Omega} =0 \quad \text{and} \quad A^{(1)}_r=0 \,.
\end{equation}
The background $g_{(0)}$ exhibits a residual $\text{SO}(p)$ invariance. We can use this to split the system up into sectors of $\text{SO}(p)$. The scalar sector contains 4 scalars, $A_v^{(1)}$, $g_{vr}^{(1)}$, $g_{vv}^{(1)}$ and $\text{Tr}g^{(1)}_{ij}$. The vector sector contains
3 vectors $A^{(1)}_i$, $g_{vi}^{(1)}$ and $g_{ri}^{(1)}$. Finally, the tensor sector contains 1 tensor $\overline{g}_{ij}^{(1)}\equiv g_{ij}^{(1)}-\frac{1}{p}(\text{Tr}g^{(1)}_{kl})\delta_{ij}$ (the traceless part of $g_{ij}^{(1)}$).\newpage\noindent We parameterize the three $\text{SO}(p)$ sectors according to
\begin{equation*} 
\textbf{Scalar:} \ A^{(1)}_v = -\sqrt{N\gamma_0(1+\gamma_0)} \thinspace \frac{r_0^n}{r^n} h^{-1} a_v, \ \ g_{vr}^{(1)}=h^{B-N/2}f_{vr}, \ \ g_{vv}^{(1)}=h^{-1}f_{vv}, \ \ \text{Tr} \thinspace g^{(1)}_{ij}=h^B \text{Tr}f_{ij} \,,
\end{equation*}
\begin{equation} \label{eqn:decom}
\textbf{Vector:} \ A^{(1)}_i= -\sqrt{ N \gamma_0(1+\gamma_0)} \thinspace a_i, \ \ g_{vi}^{(1)}=h^{B}f_{vi}, \ \ g_{ri}^{(1)}=h^{B-N/2}f_{ri} \,,
\end{equation}
\begin{equation*}
\textbf{Tensor:} \ \overline{g}_{ij}^{(1)}=h^B \overline{f}_{ij} \,,
\end{equation*}
where $ \overline{f}_{ij}\equiv f_{ij}-\frac{1}{p}(\text{Tr}f_{kl}) \delta_{ij}$. The parameterization is chosen in such a way that the resulting EOMs only contain derivatives of $f_{ab}$ and $a_a$ and will thus be directly integrable.

\subsection{A digression: Reduction of Einstein-Maxwell theory}
\label{sec:redEM}

Here we explain how it is possible to treat general $n$ and $p$ by integrating out the transverse non-fluid dynamic directions.

In order to work out the full set of solutions and find the general form of the stress tensor and current, it is enough to consider fluid dynamic fluctuations in $1+d$ ($2\leq d<p$) directions of the brane.\footnote{We thank Joan Camps for pointing this out.} In particular, it is enough to consider $d=2$. Indeed, since the background is $\text{SO}(p)$ invariant, the correction $\text{d}s^2_{(1)}$ will consist of $\text{SO}(p)$ invariant tensor structures. The same holds for the effective blackfold stress tensor and current. In order to identify these tensor structures, it is enough to consider fluctuations in only $1+d$ directions (time + $d$ flat spatial directions) of the brane. 
Considering only fluctuations in $1+d$ brane dimensions, the metric is 
of the form (reduction of the $p$-brane with $n+2$ transverse dimensions)
\begin{equation}
\text{d}s^2=\text{d}s_{(f)}^2+e^{2\psi(\sigma_f)}\text{d}\Omega_{(n+1)}^2+e^{2\phi(\sigma_f)}\sum_{i=d+1}^{p}(\text{d}\sigma^i)^2 \,,
\end{equation}
with the one-form gauge field of the form $A_\mu=A_a(\sigma_f)$. Here the subscript $f$ means 'fluid' since the $d+2$-dimensional base space with the metric 
$\text{d}s_{(f)}^2$ will contain the fluid dynamical degrees of freedom in our computations. Integrating out the $S^{n+1}$ and $\mathbb{T}^{p-d}$ (see appendix \ref{sec::reduction}), the EOMs of the 
system take the form 
\begin{equation} \label{eqn:redeqn}
\begin{split}
R_{ab}^{(f)}&=\mathcal F_{ab}+(n+1)\left(\nabla_a \psi\nabla_b \psi+\nabla_a\nabla_b \psi\right)+(p-d)\left(\nabla_a \phi\nabla_b \phi+\nabla_a\nabla_b \phi\right) \,, \\
\square \psi &+ \left[(p-d)\nabla_b\phi+(n+1)\nabla_b\psi \right]\nabla^b \psi  =ne^{-2\psi}+\kappa \,, \\
\square \phi &+ \left[(p-d)\nabla_b\phi+(n+1)\nabla_b\psi \right]\nabla^b \phi  =\kappa \,, \\
\nabla_a F^{ab}&=j^b \,,
\end{split}
\end{equation}
where the tensor $\mathcal F_{ab}$, vector $j^a$, and scalar $\kappa$ are given by
\begin{equation}
\mathcal F^a_{\ b}=\frac{1}{2}F^{ac}F_{bc}-\kappa \delta^a_{\ b}, \ j^a=F^{ab}\left((n+1)\nabla_b \psi+(p-d)\nabla_b \phi)\right), \ \kappa=\frac{F_{ab}F^{ab}}{4(p+n+1)} \,.
\end{equation}
Working with these effective EOMs allows us to treat a general number of transverse and brane dimensions.

\section{First order equations} \label{sec:firstorder}

In order to compute the effective stress tensor and current and thereby extract the transport coefficients, we need the large $r$ asymptotics of the perturbation functions which are decomposed and parametrized according to equation (\ref{eqn:decom}). We denote the first order Einstein and Maxwell equations by
\begin{equation}
\begin{split}
	R_{\mu\nu} - \frac{1}{2} F_{\mu \rho} F_{\nu}^{\phantom{\nu}\rho} + \frac{1}{4(n+p+1)} F_{\rho\sigma}F^{\rho\sigma} g_{\mu\nu} &\equiv \varepsilon \mathcal{E}_{\mu\nu} +\mathcal{O}(\varepsilon^2) = 0 \,, \\
	\nabla_{\rho} F^{\rho}_{\phantom{\rho}\mu}  &\equiv \varepsilon \mathcal{M}_{\mu} +\mathcal{O}(\varepsilon^2) = 0  \,.
\end{split}
\end{equation}
In this section we will find the solution to each $\text{SO}(p)$ sector in turn and explain how the regularity on the horizon is ensured.

\subsection{Scalars of \texorpdfstring{$SO(p)$}{SO(p)}}


The scalar sector consists of seven independent equations which correspond to the vanishing of the components: $\mathcal{E}_{vv}, \mathcal{E}_{rv}, \mathcal{E}_{rr}, \text{Tr} \mathcal{E}_{ij}, \mathcal{E}_{\Omega\Omega}, \mathcal{M}_v, \mathcal{M}_r$.\footnote{In the reduction scheme outlined in section \ref{sec:redEM} we have $\mathcal{E}_{\Omega\Omega} = \mathcal{E}_{\psi}$, where $\mathcal{E}_{\psi}$ is the EOM for $\psi$ given in \eqref{eqn:redeqn}. Similarly, we have $\text{Tr} \mathcal{E}_{ij} = \text{Tr} \mathcal{E}^{(f)}_{ij} - (p - d) h^B \mathcal{E}_{\phi}$, where $\mathcal{E}_{\phi}$ is the EOM for $\phi$.}

\paragraph{Constraint equations:}

There are two constraint equations; $\mathcal{E}^r_{\phantom{r}v}=0$ and $\mathcal{M}_r=0$. The two equations are solved consistently by
\begin{equation}
	\left( n + 1 + pB\gamma_0 \right) \partial_v r_0 = -r_0 ( 1 - B\gamma_0 )  \partial_i u^i  \,,
\end{equation}
and
\begin{equation}
	\left( n + 1 + pB\gamma_0 \right) \partial_v \gamma_0 = -2 \gamma_0(1+\gamma_0) \partial_i u^i  \,.
\end{equation}
The first equation corresponds to conservation of energy while the second equation can be interpreted as current conservation. These are equivalent to the scalar conservation equations given by \eqref{eqn:ConsEqns} in the rest frame.

We now proceed to solve for the first order correction to the scalar part of the metric and gauge field under the assumption that the fluid configuration satisfy the above constraints. Imposing the constraint equations will make $\mathcal{E}_{rv}$ and $\mathcal{E}_{vv}$ linear related and one is therefore left with five equations with four unknowns.

\paragraph{Dynamical equations:}

The coupled system constituted by the dynamical equations is quite intractable. One approach to obtaining the solution to the system is to decouple the trace function $\text{Tr} f_{ij}$. Once $\text{Tr} f_{ij}$ is known, it turns out, as will be presented below, all the other functions can be obtained while ensuring that they are regular on the horizon.

It is possible to obtain a 3rd order ODE for $\text{Tr} f_{ij}$ by decoupling it through a number of steps. One way is to use $\mathcal{E}_{rr}$ to eliminate $f'_{rv}$ and then take linear combinations of the remaining equations. The resulting combinations can then be used to eliminate $f'_{vv}$ and $f''_{vv}$ such that one is left with two equations in terms of $a_v$ and $\text{Tr} f_{ij}$ which can then be decoupled by standard means. The resulting equation is schematically of the form
\begin{equation}
	H^{(n,p)}_3(r) \left[ \text{Tr} f_{ij} \right] '''(r) + H^{(n,p)}_2(r) \left[ \text{Tr} f_{ij} \right]''(r) + H^{(n,p)}_1(r) \left[ \text{Tr} f_{ij} \right]'(r) = S_{\text{Tr}}(r) \,,
	\label{eqn:dseqn}
\end{equation}
where $H_1, H_2$ and $H_3$ do not depend on the sources (world-volume derivatives) and the source term $S_{\text{Tr}}$ only depends on the scalar $\partial_i u^i$. The expressions for these functions are however very long and have therefore been omitted. After some work, one finds that the equation is solved by
\begin{equation} \label{eqn:tracesol}
	\text{Tr} f_{ij}(r) = c^{(1)}_{\text{Tr}} + \gamma_0 c^{(2)}_{\text{Tr}} G(r)  -  2(\partial_{i}u^{i}) \text{Tr} f_{ij}^{\text{(s)}}(r) \,,
\end{equation}
where the terms containing the two integration constants $c^{(1)}_{\text{Tr}}$ and $c^{(2)}_{\text{Tr}}$ correspond to the homogeneous solution. The entire family of homogeneous solutions to equation \eqref{eqn:dseqn} of course has an additional one-parameter freedom which has been absorbed in the particular solution $\text{Tr} f_{ij}^{\text{(s)}}(r)$ and been used to ensure horizon regularity\footnote{Note that equation \eqref{eqn:dseqn} has been derived under the assumption that $\partial_i u^i \neq 0$. This especially means that when there are no sources the one-parameter freedom disappears in accordance with \eqref{eqn:tracesol}.}. The function $G$ is given by
\begin{equation}
	G(r) = p \, \frac{r_0^n}{r^n} \left( 2 + pB \frac{r_0^n}{r^n} \gamma_0  \right)^{-1} \,.
\end{equation}
The particular solution which is regular on the horizon is given by
\begin{equation}
	\text{Tr} f_{ij}^{\text{(s)}}(r) = - \frac{r_0}{n}  (1+\gamma_0)^{\frac{N}{2}} \alpha \gamma_0 G(r) + \left( r_{\star} - \frac{r_0}{n} \left( 1 + \gamma_0 \right)^{\frac{N}{2}} \log f(r) \right) \left( 1 - \beta \gamma_0  G(r) \right) \,,
\end{equation}
with the coefficients
\begin{equation}
	\alpha =  2B \left[ \frac{2(n+1)+pB\gamma_0}{(n+1 + pB\gamma_0)^2} \right] \quad \text{and} \quad \beta = B \left[ \frac{n+2+pB\gamma_0}{n+1+ pB\gamma_0} \right]  \,.
\end{equation}
With $\text{Tr} f_{ij}$ given, the equation $\mathcal{E}_{rr}=0$ will provide the derivative of $f_{rv}$,
\begin{equation} 
	f'_{rv}(r) = \frac{r}{\left(2(n+1) + pB \frac{r_0^n}{r^n} \gamma_0 \right) h(r)^{\frac{B}{2}}} \frac{d}{dr}\left[ h(r)^{\frac{N}{2}} [\text{Tr} f_{ij}]'(r) \right]  \,. \label{eqn:dfrv}
\end{equation}
Since the equation is a 1st order ODE, the regularity of the horizon is ensured by $\text{Tr} f_{ij}$. Note that it is possible to perform integration by parts and use that the derivative of $r_{\star}$ takes a simpler form. One can thereafter obtain an analytical expression for the resulting integral. This expression is however rather long and does not add much to the question we are addressing for which we are in principle only interested in the large $r$ behavior given by
\begin{equation}
	f_{rv}(r) \approx c_{rv} - \gamma_0 c^{(2)}_{\text{Tr}}  \frac{ G(r) h(r)}{ \left( 2 + pB\frac{r_0^n}{r^n} \gamma_0 \right)} + (\partial_i u^i) \sum_{k=1}^{\infty} \frac{r_0^{nk}}{r^{nk}} \left[  \alpha^{(k)}_{rv} r + \beta^{(k)}_{rv} r_0  \right]  \,.
\end{equation}
The first two terms constitute the homogeneous solution and the particular solution is given in terms of the coefficients $\alpha^{(k)}_{rv}$ and $\beta^{(k)}_{rv}$ which depend on $n,p$, and $\gamma_0$. The first set of coefficients are given in appendix \ref{sec:sCoeff}. \vspace{5 mm}

Using the expression for $f'_{rv}$ in terms of $\text{Tr} f_{ij}$, the Maxwell equation $\mathcal{M}_v=0$ becomes a 2nd order ODE for the gauge field perturbation,
\begin{equation} 
	\frac{d}{dr}\left[ \frac{1}{r^{n-1}} a'_v(r) \right] = \frac{n r^2}{ \left(2(n+1) + pB \frac{r_0^n}{r^n} \gamma_0 \right)} \frac{d}{dr}\left[ \frac{1}{r^{n+1}} [\text{Tr} f_{ij}]'(r) \right] \,. \label{eqn::av}
\end{equation}
This equation is solved by a double integration. The inner integral is manifestly regular at the horizon, one can therefore work directly with the asymptotic behavior of the right-hand side before performing the integrations. The large $r$ behavior of the perturbation function is thus found to be
\begin{equation}
	a_{v}(r) \approx c_v^{(1)} r^n + c_v^{(2)} + \frac{1}{2} \gamma_0 c^{(2)}_{\text{Tr}} G(r) + (\partial_i u^i) \left[ - \frac{n}{n-1} r + \sum_{k=1}^{\infty} \frac{r_0^{nk}}{r^{nk}} \left[ \alpha_v^{(k)} r + \beta_v^{(k)} r_0 \right] \right] \,,
\end{equation}
where the first three terms constitute the homogeneous solution and the particular solution is given in terms of the coefficients $\alpha^{(k)}_{v}$ and $\beta^{(k)}_{v}$ depending on $n,p$, and $\gamma_0$. The first set of coefficients are given in appendix \ref{sec:sCoeff}. \vspace{5 mm}

The last perturbation function $f_{vv}$ can be obtained from $\mathcal{E}_{vv}=0$. Using the expression for $f'_{rv}$ in terms of $\text{Tr} f_{ij}$ the equation is schematically of the form
\begin{equation}
	\frac{d}{dr}\left[ r^{n+1} f'_{vv}(r) \right] = G_1\left[ \text{Tr} f_{ij}(r) \right] + G_2\left[a_v(r) \right]  + S_{ii}(r)  \,, \label{eqn:fvv}	
\end{equation}
where $G_1, G_2$ are non-trivial differential operators and the source $S_{ii}$ depends on $\partial_i u^i$. Again, the full expressions have been omitted and we only provide the large $r$ behavior,
\begin{equation}
	f_{vv}(r) \approx f^{\text{(h)}}_{vv}(r) + (\partial_i u^i)
	\sum_{k=1}^{\infty} \frac{r_0^{nk}}{r^{nk}} \left[ \alpha_{vv}^{(k)} r + \beta_{vv}^{(k)} r_0 \right]	\,,
\end{equation}
with the homogeneous part given by
\begin{equation}
	f^{\text{(h)}}_{vv}(r) = c_{vv}^{(1)} + \frac{c_{vv}^{(2)}}{r^n} + \frac{r_0^{n}}{r^{n}} \frac{\gamma_0}{h(r)} \left[ 2(1+\gamma_0) \frac{r_0^{n}}{r^{n}} (c_v^{(2)} - c_v^{(1)}r_0^n \gamma_0) - \frac{G(r) }{2} \left( 1 + 2\gamma_0 - \frac{r_0^n}{r^n}\gamma_0 \right) c^{(2)}_{\text{Tr}} \right] \,.
\end{equation}
The solution is ensured to be regular at the horizon. The coefficients $\alpha^{(k)}_{vv}$ and $\beta^{(k)}_{vv}$ depend on $n,p$, and $\gamma_0$. The first set of coefficients are listed in appendix \ref{sec:sCoeff}.

Finally, one must ensure that the equations coming from $\text{Tr} \mathcal{E}_{ii}$ and the angular directions ($\mathcal{E}_{\Omega\Omega}=0$) are satisfied. This will impose the following relations between the integration constants,
\begin{eqnarray}
	c^{(1)}_{vv} &=& - 2 c_{rv} \,, \\
	c^{(2)}_{vv} &=& - \frac{r_0^n}{4} \left( (n+p)c^{(2)}_{\text{Tr}} + 8(1+\gamma_0)(c_v^{(2)} - c_v^{(1)}r_0^n \gamma_0) \right)  \,. \label{eqn:OOcons}
\end{eqnarray}
This completes the analysis of the scalar sector. The remaining undetermined integration constants are thus: $c^{(1)}_{\text{Tr}}, c^{(2)}_{\text{Tr}}, c_{rv}, c_v^{(1)}, c_v^{(2)}$ for which $c_{rv}$ and $c^{(1)}_{\text{Tr}}$ will be fixed by requiring the spacetime to be asymptotically flat while the rest constitute the freedom of the homogeneous solution. Note that the above functions reproduce the neutral case as $\gamma_0(\sigma^a) \rightarrow 0$.

\subsection{Vectors of \texorpdfstring{$SO(p)$}{SO(p)}}

The vector sector consists of $3p$ independent equations which correspond to the vanishing of the components: $\mathcal{E}_{ri}, \mathcal{E}_{vi}$ and $\mathcal{M}_i$.

\paragraph{Constraint equations:}

The constraint equations are given by the Einstein equations $\mathcal{E}^r_{\phantom{r} i}=0$ and are solved by
\begin{equation} \label{eqn:consVecSec}
	\partial_i r_0 = r_0 (1 + N \gamma_0) \partial_v u_i \,,
\end{equation}
which are equivalent to conservation of stress-momentum. These are part of the conservation equations given by \eqref{eqn:ConsEqns} in the rest frame. Similar to above we now proceed solving for the first order corrections to the metric and gauge field under the assumption that the fluid profile satisfy the above constraint \eqref{eqn:consVecSec}.

\paragraph{Dynamical equations:}

The remaining equations consist of $p$ pairs consisting of one Einstein equation $\mathcal{E}_{v i}=0$ and one Maxwell equation $\mathcal{M}_{i}=0$. The structure of these equations is the same as in the scalar sector. The Einstein equation $\mathcal{E}_{vi}=0$ is schematically of the form,
\begin{equation}
	L_3^{(n,p)}(r) f_{vi}''(r) + L_2^{(n,p)}(r) f_{vi}'(r) + L_1^{(n,p)}(r) a_i'(r)  = S_{vi}(r) \,,
\end{equation}
while the Maxwell equation $\mathcal{M}_{i}=0$ is,
\begin{equation}
	M_3^{(n,p)}(r) a_i''(r) + M_2^{(n,p)}(r) a_i'(r) + M_1^{(n,p)}(r) f_{vi}'(r) = S_{i}(r) \,.
\end{equation}
Again the functions $L_k$ and $M_k$, $k=1,\ldots,3$ have been omitted.

To decouple the system we differentiate $\mathcal{E}_{vi}$ once and eliminate all $a_i(r)$ terms in $\mathcal{M}_{i}$. Doing so, one obtains a 3rd order ODE for $f_{vi}(r)$ which can be written on the form,
\begin{equation}
	\frac{d}{dr} \left[ \frac{r^{n+1} f(r)}{h^{N}} \left( 1 - c_1 \frac{r_0^n}{r^n} \right)^2 \frac{d}{dr} \left[ \frac{ r^{n+1} h^{N+1}}{ \left( 1 - c_1 \frac{r_0^n}{r^n} \right) } f'_{vi}(r) \right] \right] 
	= S_{vi}(r) \,,
\end{equation}
with
\begin{equation}
	c_1 = \frac{N-1}{1+ N\gamma_0}\gamma_0 \,.
\end{equation}
It is possible to perform the first two integrations analytically and ensure regularity at the horizon. The first integration is straightforward while the second involves several non-trivial functions. The large $r$ behavior of the $f_{vi}$ function is found to be
\begin{equation}
	f_{vi}(r) \approx c^{(1)}_{vi} - \left( 1- \frac{f(r)}{h(r)^N} \right) c_{vi}^{(2)} - (\partial_v u_i) r + \sum_{k=1}^{\infty} \frac{r_0^{nk}}{r^{nk}} \left[ \alpha^{(k)}_{vi} r  + \beta^{(k)}_{vi} r_0 \right]  \,,
\end{equation}
where the first two terms constitute the homogeneous solution. The first set of coefficients $\alpha^{(k)}_{vi}$ and $\beta^{(k)}_{vi}$ are given in appendix \ref{sec:sCoeff}. Notice that the sum vanishes in the neutral limit. \vspace{5mm}

Once the solution of $f_{vi}$ is given we can use $\mathcal{E}_{vi}$ to determine $a_i$,
\begin{equation}
	a_{i}(r) \approx  c^{(1)}_{i} + \frac{r_0^n}{r^n}\frac{1}{h(r)} c_{vi}^{(2)} + \sum_{k=1}^{\infty} \frac{r_0^{nk}}{r^{nk}} \left[ \alpha^{(k)}_{i} r  + \beta^{(k)}_{i} r_0 \right]\,,
\end{equation}
where the first two terms correspond to the homogeneous solution. The first set of coefficients $\alpha^{(k)}_{i}$ and $\beta^{(k)}_{i}$ are given in appendix \ref{sec:sCoeff}.

The remaining undetermined integration constants are thus: $c^{(1)}_{i}, c^{(1)}_{vi}$, and $c^{(2)}_{vi}$. The constant $c^{(2)}_{vi}$ corresponds to an infinitesimal shift in the boost velocities along the spatial directions of the brane while $c^{(1)}_{i}$ is equivalent to an infinitesimal gauge transformation. The last constant $c^{(1)}_{vi}$ will be determined by imposing asymptotically flatness at infinity.

\subsection{Tensors of \texorpdfstring{$SO(p)$}{SO(p)}}

There are no constraint equations in the tensor sector and $p(p+1)/2-1$ dynamical equations given by
\begin{equation}
	\mathcal{E}_{ij} - \frac{\delta_{ij}}{p} \text{Tr}(\mathcal{E}_{ij}) = 0 \,.
\end{equation}
This gives an equation for each component of the traceless symmetric perturbation functions $\bar{f}_{ij}$,
\begin{equation}
	\frac{d}{dr} \left[ r^{n+1} f(r) \bar{f}'_{ij}(r) \right] = - \sigma_{ij} r^n \left( 2(n+1) + pB \frac{r_0^n}{r^n} \gamma_0 \right) h(r)^{\frac{B}{2}} \,,
\end{equation}
where
\begin{equation}
	\sigma_{ij} = \partial_{(i}u_{j)} - \frac{1}{p}\delta_{ij} \partial_k u^k \,.
\end{equation}
The solution is given by,
\begin{equation}
	\bar{f}_{ij}(r) = \bar{c}_{ij} - 2\sigma_{ij} \left( r_{\star} - \frac{r_0}{n} \left( 1 + \gamma_0 \right)^{\frac{N}{2}} \log f(r) \right) \,,
\end{equation}
where horizon regularity has been imposed and the constant $\bar{c}_{ij}$ is symmetric and traceless and will be determined by imposing asymptotically flatness.

\subsection{Comment on the homogeneous solution}

We have now obtained the solution to the Einstein-Maxwell equations for any first order fluid profile which fulfill the constraint equations. These have been provided in large $r$ expansions and are ensured to have the right behavior at the horizon for any of the remaining integration constants. One remark that is worth mentioning is that $f_{ri}$ did not appear in the analysis above and corresponds to a gauge freedom. This gauge freedom does not play a role for $n\geq2$, but is expected to play a role for $n=1$ to ensure asymptotically flatness.

We now want to provide some insight into the meaning of the remaining integration constants. One can separate the constants into two categories; the subset that are fixed by asymptotically flatness and the subset that corresponds to the $\varepsilon$-freedom of the parameters in the zeroth order fields. The latter corresponds exactly to the remaining freedom of the homogeneous solution. In the above the homogeneous part of the fields are given exact.

One finds that the homogeneous part of the scalar sector corresponds to shifts in $r_0 \rightarrow r_0 + \varepsilon \delta r_0$, $\gamma_0 \rightarrow \gamma_0 + \varepsilon \delta \gamma_0$, and the gauge freedom $a_v \rightarrow a_v + \varepsilon \delta a_v $ of the zeroth order metric given by equation \eqref{eqn:leadGMmetric}. Indeed, performing the above shifts and redefining the $r$ coordinate,
\begin{eqnarray}
	r &\rightarrow& r \left( 1 -  \varepsilon \gamma_0 \frac{ r_0^{n}}{r^n} \frac{n \delta \log r_0 + \delta \log \gamma_0}{  n + p h(r)}  \right) \,,
\end{eqnarray}
such that the angular directions does not receive first order contributions in accordance with the gauge choice \eqref{eqn:gaugechoice2}, one can relate the integration constants to the two shifts and gauge transformation by,
\begin{eqnarray} \label{eqn:relations}
	c^{(2)}_{\text{Tr}} &=& 2B \left(n \delta \log r_0 + \delta \log \gamma_0 \right) \,, \nonumber \\
	c^{(2)}_v &=& -n\delta \log r_0  - \frac{1+2\gamma_0}{2(1+\gamma_0)} \delta \log \gamma_0 - \frac{\gamma_0}{\sqrt{N \gamma_0 (1+\gamma_0)}} \delta a_v \,, \\
	c^{(1)}_v &=& -\frac{\delta a_v}{r_0^n \sqrt{N \gamma_0 (1+\gamma_0)}} \nonumber \,.
\end{eqnarray}

For the vector sector one finds that the homogeneous part corresponds to the shift of $u_i \rightarrow u_i + \varepsilon \delta u_i$ and the gauge transformation $a_i \rightarrow a_i + \varepsilon \delta a_i$. The first transformation corresponds to global shifts in the boost velocities. In the same $r$-coordinate, one has
\begin{eqnarray}
	c^{(2)}_{vi} &=& \delta u_i \,, \nonumber \\
	c^{(1)}_{i} &=& -\frac{\delta a_i}{\sqrt{N \gamma_0 (1+\gamma_0)}} \,.
\end{eqnarray}
This accounts for all the $\varepsilon$-freedom in the full solution.

\subsection{Imposing asymptotically flatness}

We now turn to imposing the boundary condition at infinity, namely requiring the solution to be asymptotically flat. To impose this we must first change coordinates back to the Schwarzschild-like form. Moreover, we need the fields expressed in Schwarzschild coordinates for obtaining the effective stress tensor and current. In order to change coordinates, we use the inverse transformation of the one stated in equation (\ref{eqn:EFcoord}). The transformation can be worked out iteratively order by order. To first order the transformation from EF-like coordinates to Schwarzschild-like coordinates for a general $r_0(\sigma^a)$ and $\gamma_0(\sigma^a)$ is given by,
\begin{equation}
\begin{split}
	v &= t + r_{\star} + \varepsilon \bigg[ (t + r_{\star}) \left( \partial_{r_0} r_{\star} \partial_t r_0 + \partial_{\gamma_0} r_{\star} \partial_t \gamma_0  \right) +  x^i \left( \partial_{r_0} r_{\star} \partial_i r_0 + \partial_{\gamma_0} r_{\star} \partial_i \gamma_0  \right) \bigg] + \mathcal{O}(\varepsilon^2) \,, \\
	\sigma^i &= x^i + \varepsilon \bigg[ (t+r_{\star}) \partial_t u^i + \sigma^j \partial_j u^i \bigg] r_{\star} + \mathcal{O}(\varepsilon^2) \,.
\end{split}
\end{equation}
It is now possible to transform all the fields to Schwarzschild coordinates and impose asymptotically flatness. This leads to
\begin{eqnarray}
	c_{rv} = 0, \quad c^{(1)}_{vi} = 0, \quad c^{(1)}_{\text{Tr}} = 0, \quad \bar{c}_{ij} = 0 \,.
\end{eqnarray}
We now have the complete first order solution for the black brane metric and Maxwell gauge field that solves the Einstein-Maxwell equations.

\section{Viscous stress tensor and current} \label{sec:stresstensor}

In this section we will compute the effective stress tensor and current of the first order solution obtained above. Before doing this, we shall briefly discuss the general form of the first order derivative corrections to the stress tensor and current.

\subsection{First order fluid dynamics} \label{sec:fluidanalysis}

We write  the stress tensor and the current as 
\begin{equation}
T^{ab}=T^{ab}_{(0)}+\Pi^{ab}_{(1)}+\mathcal{O}(\partial^2), \quad J^{a}=J^{a}_{(0)}+\Upsilon^{a}_{(1)}+\mathcal{O}(\partial^2) \,,
\end{equation} 
where the perfect fluid terms were written down for our specific fluid in section \ref{sec:termosec}. The tensors 
$\Pi^{ab}_{(1)}$ and $\Upsilon^{ab}_{(1)}$ are the first order dissipative derivative corrections to the perfect fluid stress tensor and current, respectively. The specific form of $\Pi^{ab}_{(1)}$ and $\Upsilon^{ab}_{(1)}$ are encoded in the first order 
correcting solution obtained in the previous section. As is well-known, to any order in derivatives, it is in principle possible to write down all the terms that can contribute to the stress tensor and current (see e.g. \cite{Romatschke:2009kr}). In this way the dissipative corrections to the stress tensor and the current can be characterized in terms of a set of transport coefficients. It is possible to show that the most general form of $\Pi^{ab}_{(1)}$ is given by\footnote{Here we have imposed the Landau frame gauge on the stress tensor $u_a \Pi^{ab}_{(1)}=0$. Similarly one can impose a Landau frame condition on the current. It takes the form $u_a \Upsilon^{a}_{(1)}=0$.}
\begin{equation}
\Pi^{ab}_{(1)}=-2\eta \sigma^{ab}-\zeta \vartheta \Delta^{ab} \,,
\end{equation}
where $\sigma^{ab}$ is the usual shear tensor and is given by
\begin{equation} \label{eqn:firstorderstress}
\sigma^{ab} = \Delta^{ac}\left(\partial_{(c}u_{d)}-\Delta_{cd}\frac{\vartheta}{p}\right) \Delta^{db} \quad \text{with} \quad \vartheta = \partial_a u^a \,,
\end{equation}
The coefficients $\eta$ and $\zeta$
are respectively the shear and bulk viscosity transport coefficients and were computed for the neutral brane in \cite{Camps:2010br}. The bulk and shear viscosities are associated with the scalar and tensor fluctuations, respectively. Note that although the overall form of $\Pi_{(1)}^{ab}$ is the same as in the neutral case,
the transport coefficients are now expected to depend on both the temperature and the charge i.e. on both $r_0$ and $\gamma_0$. Also note that the viscosities $\eta$ and $\zeta$ are required to be positive in order to ensure entropy creation in the fluid \cite{landau_fluid_1987}. 

Using similar reasoning, it is possible to show that the most general form of $\Upsilon^{a}_{(1)}$ (in the Landau frame) is given by\footnote{It is possible to include a parity violating term as was found in \cite{Erdmenger:2008rm}. However, since we have no Chern-Simons term in the theory such a term is not relevant.}
\begin{equation}\label{currentcorr}
\Upsilon^{a}_{(1)}= - \mathfrak D \left( \frac{\mathcal{Q} \mathcal{T}}{w} \right)^2  \Delta^{ab}\partial_b\left(\frac{\Phi}{\mathcal T}\right) \,.
\end{equation}
Here $\mathfrak D$ is the charge diffusion constant which is associated with the vector fluctuations. Indeed, it is possible to derive that with $\mathfrak D>0$, the term \eqref{currentcorr} is the only term which can be constructed from the fields and that is consistent with the 2nd law 
of thermodynamics \cite{landau_fluid_1987}. Plugging in the values of $\Phi$ and $\mathcal T$ in terms of $r_0$ and $\gamma_0$ and using the vector constraint equation \eqref{eqn:consVecSec}, we find that (in the rest frame)
\begin{equation} \label{eqn:proptodiff}
\Upsilon^{v}_{(1)}=0, \quad \Upsilon^{i}_{(1)}\sim \gamma_0(1+\gamma_0)\partial_v u^i+\frac{1}{2}\partial_i \gamma_0 \,.
\end{equation} 
Since the derivatives appear in a very specific combination in this expression, this in fact provides us with a non-trivial check of the blackfold fluid description.

\subsection{Computing the effective stress tensor and current} 

The quasi-local stress tensor $\tau_{\mu\nu}$ is obtained by background subtraction. We consider a surface at large $r$ (spatial infinity) with induced metric $h_{\mu\nu}$ and inwards pointing normal vector and compute the components of the quasi-local tensor by,
\begin{equation}
	8\pi G \tau_{\mu\nu} = K_{\mu\nu} - h_{\mu\nu} K - \left( \hat{K}_{\mu\nu} - h_{\mu\nu} \hat{K} \right) \,,
\end{equation}
where $K_{\mu\nu}$ is the extrinsic curvature of the surface and $K=g^{\mu\nu}K_{\mu\nu}$. The hatted quantities are the subtracted terms which are computed on flat spacetime with the same intrinsic geometry on both boundaries. Notice that the transverse space bear the structure $h^B d\Omega^2_{(n+1)}$. One finds that for the transverse directions $\tau_{\Omega\Omega} = 0$ while for the brane directions we obtain the fluid stress tensor by,
\begin{equation}
	T_{ab} = \lim_{r \rightarrow \infty} \frac{\Omega_{(n+1)} }{2} r^{n+1} \tau_{ab} \,,
\end{equation}
where $\Omega_{(n+1)}$ is the volume of the $(n+1)$-sphere. We find
\begin{equation} 
\begin{split}
T_{tt} &= \frac{\Omega_{(n+1)}}{16\pi G} \left( n+1 + nN(\gamma_0 + \varepsilon ( \delta \gamma_0 + x^a \partial_a \gamma_0)) \right) \left( r_0 + \varepsilon (\delta r_0 + x^a \partial_a r_0) \right)^n  \,, \\
T_{ij} &= -\frac{\Omega_{(n+1)}}{16\pi G} \bigg( \delta_{ij} \left( r_0 + \varepsilon (\delta r_0 + x^a \partial_a r_0) \right)^n \\
& + \varepsilon r_0^{n+1} (1+\gamma_0)^{\frac{N}{2}} \left[ 2\left( \partial_{(i}u_{j)} - \frac{1}{p}\delta_{ij} \partial_k u^k \right) + \frac{2}{p} \frac{(n+p+1)(n+1)}{\left( n+1 + pB \gamma_0 \right)^2} \delta_{ij} \partial_k u^k \right] \bigg) \,,  \\
T_{tj} &= -  \frac{\Omega_{(n+1)}}{16\pi G}  r_0^{n} n (1 + N \gamma_0) \varepsilon ( \delta u_j + x^a \partial_a u_j )\,,
\end{split}
\end{equation}
where the expressions are valid to order $\mathcal{O}(\varepsilon)$. In a similar manner the current is obtained from large $r$ asymptotics of the gauge fields. Ensuring that the Lorenz gauge condition $\nabla^{\mu} A_{\mu} = 0$ is satisfied, the current is obtained using
\begin{equation}
	J_{a} = \lim_{r \rightarrow \infty} \frac{n \Omega_{(n+1)} }{16 \pi G} r^n A_{a} \,.
\end{equation}
One finds
\begin{align}
\begin{split}
	J_{t} &= -\frac{\Omega_{(n+1)}}{16\pi G} n \sqrt{N} \left( r_0 + \varepsilon (\delta r_0 + x^a \partial_a r_0) \right)^n \sqrt{ \gamma_0(1+\gamma_0) + \varepsilon(\delta \gamma_0 + x^a \partial_a \gamma_0)(1 + 2 \gamma_0) } \,, \\
	J_{i} &= \frac{\Omega_{(n+1)}}{16\pi G} n \sqrt{N} r_0^n \sqrt{\gamma_0(1+\gamma_0)}  \Bigg( \varepsilon \big( \delta u_j + x^a \partial_a u_j \big) 
	 - \varepsilon r_0 \frac{\gamma_0(1+\gamma_0) \partial_v u_i + \tfrac{1}{2}\partial_i \gamma_0}{ n (1 + N \gamma_0) \gamma_0 (1+\gamma_0)^{\frac{B}{2}+1} }  \Bigg) \,.
\end{split}
\end{align}
Again these expressions are valid to $\mathcal{O}(\varepsilon)$. It is now possible to read off the transport coefficients. Before doing this, we require that the Landau frame renormalization conditions $\Pi^{tt}_{(1)}=\Pi^{ti}_{(1)}=0$ and $\Upsilon^{t}_{(1)}=0$ are satisfied. Equivalently we require the shifts $\delta r_0$ and $\delta \gamma_0$ of the zeroth order solution to vanish. Notice that the stress tensor and current do not depend on the gauge transformation $\delta a_a$ as they should of course not do. Also recall that the shifts were related to the integrations constants by \eqref{eqn:relations}. 

Setting $\delta r_0= \delta \gamma_0 = 0$, the shear and bulk viscosities are determined using the form given by equation (\ref{eqn:firstorderstress}),
\begin{equation} \label{eqn:etazeta}
	\eta = \frac{\Omega_{(n+1)}}{16 \pi G} r_0^{n+1} (1+\gamma_0)^{\frac{N}{2}}, \quad \frac{\zeta}{\eta} = \frac{2}{p} \frac{(n+p+1)(n+1)}{\left( n+1 + pB \gamma_0 \right)^2} \,.
\end{equation}
The second term of $J_i$ is seen to have the right proportionality according to \eqref{eqn:proptodiff} and hence using the form of equation \eqref{currentcorr} the diffusion constant can be determined,
\begin{equation} \label{eqn:DiffusionConst}
	\mathfrak{D} = \frac{\Omega_{(n+1)}}{4 G} \frac{1+\gamma_0}{nN \gamma_0} \; r_0^{n+2} \,. 
\end{equation}
Notice that all the transport coefficients are found to be positive which is expected for a consistent effective fluid dynamic theory. We have now obtained the first order derivative corrections to the effective stress tensor and current.

\subsection{Hydrodynamic bounds}

We will now check the result of the shear viscosity against the expectation that the transport coefficient should satisfy the bound
\begin{equation}
	\frac{\eta}{s} \geq \frac{1}{4\pi} \,.
\end{equation}
Using equations \eqref{parar0gamma0} and \eqref{eqn:etazeta}, the system is seen to saturate the bound as expected. \vspace{5mm}

In addition, it is worth to investigate the bulk to shear viscosity ratio proposed by ref. \cite{Buchel:2007mf},
\begin{equation} \label{eqn:bulkbound}
	\frac{\zeta}{\eta} \geq 2\left( \frac{1}{p} - c_s^2 \right) \,
\end{equation}
where $c_s$ is the speed of sound computed below in section \ref{sec:stability}. Although one should keep in mind that the proposal of this bound relies heavily on holographic considerations, we find when using the value given by equation \eqref{eqn:speedofsound} for the Reissner-Nordstr\"om brane, that the bound is satisfied in the range
\begin{equation}
	0 \leq \gamma_0 \leq -\frac{ n+1 - \sqrt{1+n(n+p+2)} }{pB}  \,,
\end{equation}
while for large values of $\gamma_0$ the bound is found to be violated. If we instead of $c_s$ in \eqref{eqn:bulkbound} use the proposed quantity \cite{Gouteraux:2011qh, Smolic:2013gx}
\begin{equation}
	c_{\mathcal{Q}}^2 \equiv \left( \frac{\partial P}{\partial \epsilon} \right)_{\mathcal{Q}} = - \frac{1}{1+n} \left[ \frac{1 + 2\gamma_0}{1 + \frac{pB}{n+1}\gamma_0}  \right]\,,
\end{equation}
computed for fixed charge density $\mathcal{Q}$, we find that the bound will always be violated (except for the neutral case where $c_{\mathcal{Q}} = c_s$).

\section{Stability and dispersion relations} \label{sec:stability}

In ref. \cite{Emparan:2009at} the Gregory-Laflamme instability was successfully identified with the unstable sound mode of the neutral black brane. This analysis was further refined in \cite{Camps:2010br} and considered for branes charged under top-form gauge fields in \cite{Emparan:2011hg}. In this section we address the issue of stability and dispersion of long wavelength perturbations of the Reissner-Nordstr\"om black brane. Moreover, we comment on the connection to thermodynamic (in)stability.

\subsection{Dispersion relations}
It is straightforward to show that the first order fluid (conservation) equations take the form 
\begin{equation}\label{firstorderfluideqs}
\begin{split}
\dot{\epsilon}&=-(w-\zeta \vartheta)\vartheta-2\eta\sigma_{ab}\sigma^{ab} \,, \qquad
\dot{u}^a = -\frac{\Delta^{ab}\partial_b(P-\zeta\vartheta)-2\eta \Delta^{a}_{\ b}\partial_c \sigma^{bc}}{w-\zeta\vartheta}\,, \\
\dot{\mathcal Q} &= -\mathcal Q \vartheta + \mathfrak D \left( \frac{\mathcal{Q}\mathcal{T}}{w} \right)^2 \left(\vartheta u^b+\dot{u}^b + \Delta^{ab}\partial_a\right) \partial_b \left(\frac{\Phi}{\mathcal{T}}\right) \,,
\end{split}
\end{equation}
where the transport coefficients and the factor associated to $\mathfrak{D}$ are coefficients in the derivative expansion and should be treated as constants. In order to find the speed of sound and dispersion relations, we consider \emph{small} long wavelength perturbations of the fluid
\begin{equation}
\Phi\to\Phi+\delta \Phi \thinspace e^{i\left(\omega t+k_j x^j \right)}, \quad \mathcal{T}\to \mathcal{T}+\delta \mathcal{T} \thinspace e^{i\left(\omega t+k_j x^j \right)} , \quad u^a=(1,0,\dots)\to (1,\delta u^i \thinspace e^{i\left(\omega t+k_j x^j \right)}) \,.
\end{equation}
The charge density $\mathcal Q$, energy density $\epsilon$, and pressure $P$ are perturbed according to 
\begin{equation}
\mathcal Q \to \mathcal Q+\delta \mathcal Q \thinspace e^{i\left(\omega t+k_j x^j \right)}, \quad \epsilon\to \epsilon+\delta \epsilon \thinspace e^{i\left(\omega t+k_j x^j \right)} , \quad P\to P+\delta P \thinspace e^{i\left(\omega t+k_j x^j \right)} \,,
\end{equation}
where the amplitudes can be expressed in terms of thermodynamic derivatives that depend on the specific equation of state. Note that $\delta p = \mathcal{Q} \delta \Phi+s \delta T$ as a consequence of the Gibbs-Duhem relation. Plugging the expressions into the first order fluid equations \eqref{firstorderfluideqs} and linearizing in the amplitudes, we obtain the $p+2$ equations 
\begin{equation}
\begin{split}
i\omega\left(\left(\frac{\partial \epsilon}{\partial \Phi}\right)_\mathcal{T} \delta \Phi  + \left(\frac{\partial \epsilon}{\partial \mathcal{T} }\right)_\Phi \delta \mathcal{T} \right)+i w k_i \delta u^i&=0 \,, \\
iw \omega \delta u^j+i k^j \left( \mathcal{Q} \delta \Phi+s \delta \mathcal{T} \right)+k^j \left(\eta\left(1-\frac{2}{p}\right)+\zeta \right) k_i \delta u^i+\eta k^2\delta u^j&=0 \,, \\
i\omega \left( \left( \frac{\partial \mathcal{Q}}{\partial \Phi} \right)_\mathcal{T} \delta \Phi + \left(\frac{\partial \mathcal{Q}}{\partial \mathcal{T} }\right)_\Phi \delta \mathcal{T} \right)+i \mathcal Q k_i \delta u^i + 
\mathfrak{D}\mathcal{T}\frac{\mathcal{Q}^2}{w^2}\left(\delta \Phi - \frac{\Phi}{\mathcal{T}}\delta \mathcal{T} \right)k^2 &=0 \,. \end{split}
\end{equation} 
We stress that the thermodynamic derivatives are not dynamical and do only depend on the equation of state of the fluid in question. In our case they can be computed from \eqref{parar0gamma0} and \eqref{eqn:energydensity}. In order to find the $\omega$ that solves this system for a given wave vector $k^i$, we set the determinant of the system of linear equations in the amplitudes to zero. To linear order in 
$k^i$ (i.e. at the perfect fluid level) the dispersion relation gives the speed of sound $c_s=\omega/k$. Using the equation of state \eqref{eqn:energydensity} and solving the system to linear order, one finds
\begin{equation} \label{eqn:speedofsound}
c_s^2=\left(\frac{\partial P}{\partial \epsilon}\right)_{\tfrac{s}{\mathcal{Q}}}= -\frac{1-B\gamma_0}{1+N\gamma_0} \thinspace \Big(n+1+pB\gamma_0 \Big)^{-1} \,.
\end{equation}
As was found with the $p=q$ branes of supergravity \cite{Emparan:2011hg}, the speed of sound only depends on the charge parameter $\gamma_0$. For zero charge $\gamma_0=0$ we recover the neutral result $c_s^2=-1/(n+1)$. Since a negative speed of sound squared signifies an unstable sound mode, the neutral brane is unstable under long wavelength perturbations. Indeed, this instability is exactly identified
with the GL instability \cite{Camps:2010br}. However, as we increase $\gamma_0$ the speed of sound squared becomes less and less negative and for
\begin{equation}\label{chargestabcond}
\gamma_0>  \bar{\gamma}_0 = \frac{D-3}{2} \,,
\end{equation} 
the $q=0$ brane becomes stable under long wavelength perturbations to leading order. Notice that the condition \eqref{chargestabcond} can be satisfied for any non-zero charge density if the black
brane temperature is low enough. Indeed, stability is obtained for $\mathcal T \sim (G\mathcal Q)^{-1/n}$ (where the exact numerical factor depends on the number of transverse and brane dimensions). 

\begin{figure}[!t]
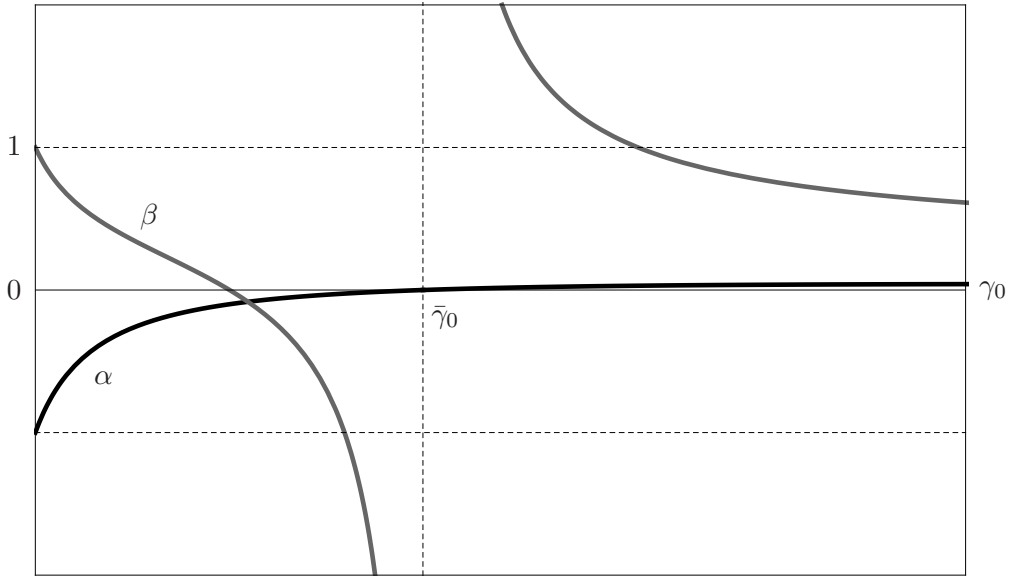

\begin{center}
\begin{lpic}{11062013soundmode(0.60,)}
\lbl[r]{-3,63;$0$}
\lbl[r]{93,57;$\bar{\gamma}_0$}
\lbl[r]{-3,95;$1$}
\lbl[t]{210,65;$\gamma_0$}
\lbl[t]{25,82;$\beta$}
\lbl[t]{15,45;$\alpha$}
\end{lpic}
\caption{The qualitative behavior of the sound mode $\omega=c_s(\gamma_0)k+a(\gamma_0)ik^2+\mathcal{O}(k^3)$ given by equation \eqref{sosdisp} as a function of $\gamma_0$. The linear term (speed of sound) is parametrized according to $c_s^2(\gamma_0) = |c_s^2(0)| \alpha(\gamma_0)$ while the quadratic term (sound mode attenuation) is parametrized as $a(\gamma_0) = |a(0)| \beta(\gamma_0)$. Note that the linear and quadratic term become positive when the charge density passes the threshold $\bar{\gamma}_0$ indicated by the vertical dashed line.}
\label{fig:soundmode}
\end{center}
\end{figure}

In order to check stability to next to leading order, we now work out the dispersion relation for the fluid to quadratic order in $k$. We solve the system of equations to $\mathcal{O}(k^2)$. Solving for the longitudinal modes, we find the equation
\begin{equation}
\omega-c_s^2\frac{k^2}{\omega}-i\frac{k^2}{w}\left(2\left(1-\frac{1}{p}\right)\eta+\zeta \right)
- \frac{ik^2}{w} \mathfrak{D} \left(  \mathcal{R}_1 \left(\frac{k}{\omega}\right)^2 + \frac{ \mathcal{R}_2 }{w} \right) +
\mathcal{O}(k^3)=0 \,,
\end{equation}  
where the coefficients $\mathcal{R}_1$, $\mathcal{R}_2$, and $\mathcal R$ (introduced below) are given in appendix \ref{sec:constR}. Solving for the sound mode(s), we find the dispersion relation 
\begin{equation}\label{sosdisp}
\omega(k)=\pm c_s k+\frac{i k^2}{w} \left(\left(1-\frac{1}{p}\right)\eta+\frac{\zeta}{2}\right)+ik^2\mathcal R \mathfrak D \,.
\end{equation}
For a general fluid both the first order term ($c_s$) and the second order term must be positive in order for it to be dynamically stable. In this case, the above equation describes dampening of the 
(long wavelength) sound waves in the fluid. Fig. \ref{fig:soundmode} shows the general behavior of $c_s$ and the (second order) attenuation term in \eqref{sosdisp}. We see that above the threshold $\bar{\gamma}_0$, the speed of sound squared and sound mode attenuation are both positive. The sound mode is therefore stable to second order.

\begin{figure}[!t]
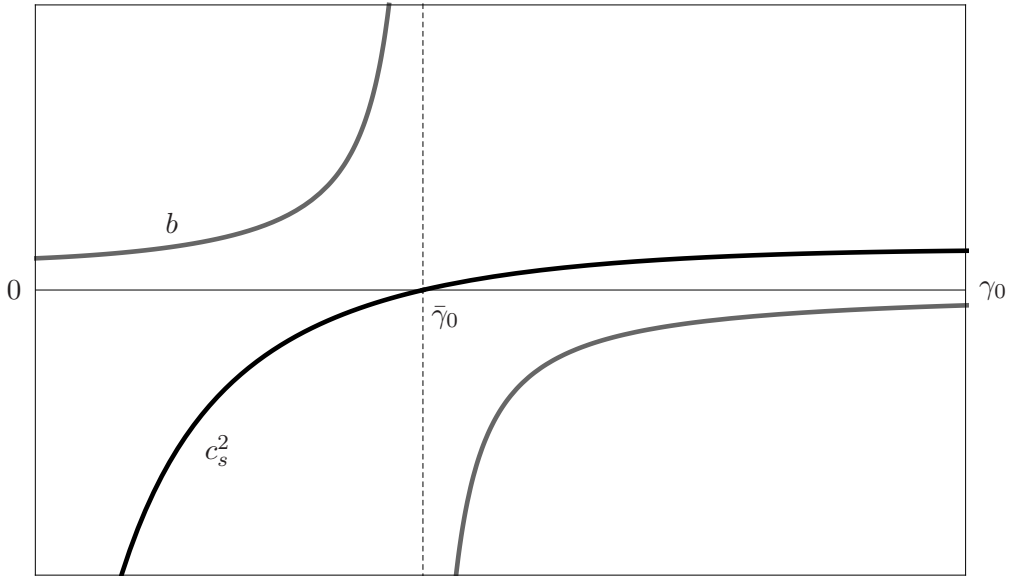

\begin{center}
\begin{lpic}{11062013chargemode(0.60,)}
\lbl[r]{-3,63;$0$}
\lbl[t]{210,65;$\gamma_0$}
\lbl[r]{93,57;$\bar{\gamma}_0$}
\lbl[t]{40,31;$c_s^2$}
\lbl[t]{30,80;$b$}
\end{lpic}
\caption{The qualitative behavior of the charge diffusion mode $\omega = i b k^2$ as a function of the charge parameter $\gamma_0$, where $b$ is given in equation \eqref{eqn:diffusionmode}. When $b$ is positive $c_s^2$ is negative and vice versa. The critical point $\bar{\gamma}_0$ is indicated by the dashed line.}
\label{fig:chargemode}
\end{center}
\end{figure}

In addition to the sound mode we have a longitudinal diffusion mode given by
\begin{equation} \label{eqn:diffusionmode}
\omega(k) = -\frac{i \mathfrak{D}\mathcal{R}_1}{c_s^2 w} k^2  =  i  \frac{(1+\gamma_0)^{1-N}}{4 \pi \mathcal{T}(1-B\gamma_0)} k^2
\,.
\end{equation}
We see that in general this mode is stable if and only if $\mathcal R_1/c_s^2<0$. In our case this amounts to the condition $\gamma_0<\bar{\gamma}_0$ i.e. the opposite of the condition \eqref{chargestabcond} as shown in Fig. \ref{fig:chargemode}. The conditions on $\gamma_0$ for dynamical stability are found to be complementary; when the sound mode is stable the charge diffusion mode is unstable and vice versa. The Reissner-Nordstr\"om brane thus seems to suffer from a GL instability for all values of the charge parameter $\gamma_0$.

Finally, we also have a shear mode which takes the form 
\begin{equation}
\omega(k)=\frac{i\eta}{w}k^2 \,.
\end{equation}
The fluctuations of the shear mode are very simple, they are transverse displacement of effective fluid with no variations in the charge and energy densities. Notice that this mode is always stable.

\subsection{Thermodynamic stability}
The conditions for thermodynamic stability of the Reissner-Nordstr\"om black brane are computed in the grand canonical ensemble since charge is allowed to redistribute itself in the directions of the brane. Using the thermodynamic quantities in equation \eqref{parar0gamma0} and \eqref{eqn:energydensity}, one finds the specific heat capacity $C$ and the (inverse) isothermal permittivity $c$ to be,
\begin{equation}
\begin{split}
  C =  \left(\frac{\partial\epsilon}{\partial\mathcal{T}} \right)_{\mathcal{Q}} &= \left( \frac{n + 1 +  (2 - n (N-2))\gamma_0}{(nN - 2)\gamma_0-1} \right) s  \,, \\
  c =  \left( \frac{\partial\Phi}{\partial\mathcal{Q}} \right)_{\mathcal{T}} &= \left( \frac{1}{(\gamma_0+1) (1 -  (nN-2)\gamma_0)} \right) \frac{1}{s\mathcal{T}}  \,.
\end{split}
\end{equation}
Thermodynamical stability is obtained if the two quantities are positive. However, these two conditions are complementary and can never be satisfied. This is also what was found for the class of smeared D$p$-branes considered in e.g. \cite{Harmark:2005jk}. Indeed, this complementary behavior is analogous to what was found for the dynamical analysis. However, the critical value of $\gamma_0$ where the quantities switch sign is not coinciding for the two analyses. It would be interesting to further investigate how the instability predicted by the dynamic analysis and the thermodynamic computation are related thus making a more precise connection to the correlated stability conjecture in the charged case \cite{Harmark:2005jk}.

\section{Discussion} \label{sec:discuss}

We have investigated the nature of the hydrodynamic effective theory that governs the intrinsic long wavelength fluctuations of the Reissner-Nordstr\"om black brane. Our analysis has extended the 
established cases of the interrelation between gravity and fluid dynamics. Although the analysis of section \ref{sec:firstorder} is quite technical, the problem at hand provides the purest example of a black brane carrying charge. With the extraction of the effective stress tensor and current, our analysis has provided the generalizations of the known neutral shear and bulk viscosities. We find that the shear viscosity receives the expected modification such that $\eta/s=1/4\pi$. Note that the entropy has the form as given in equation \eqref{parar0gamma0} for the entire family of generalized Gibbons-Maeda black branes, we therefore expect the result for $\eta$ given by \eqref{eqn:etazeta} to hold in general. In particular, this includes the case of the D3 brane. The bulk viscosity was found to be non-zero positive for all values of the charge as expected since the effective fluid is not conformal. The $\zeta/\eta$ bound proposed by ref. \cite{Buchel:2007mf} was found to be violated for certain values of the charge parameter while it was demonstrated to violate the bound proposed in \cite{Gouteraux:2011qh} in the entire range of non-zero $\gamma_0$, thus providing a counter-example. Finally, we computed the charge diffusion constant $\mathfrak D$ of the Reissner-Nordstr\"om black brane. We note that, as with the shear viscosity $\eta$, the value of $\mathfrak D$ given in \eqref{eqn:DiffusionConst} only depends on $N$ which could be an indication that the result will hold for more general cases where e.g. the black brane is charged under higher form gauge fields.

The speed of sound was found to be imaginary for small charge densities, but becomes real for sufficiently large charge parameter $\gamma_0 > (D-3)/2$. For large charge density it therefore seems that the Reissner-Nordstr\"om black brane is GL stable under long wavelength perturbations. However, including the first order corrections to the dispersion relations, one finds that the hydrodynamic mode associated with charge diffusion is unstable above the threshold value of $\gamma_0$. The Reissner-Nordstr\"om black brane is therefore GL unstable for all charge densities, although it is worth noting that the brane is ``less" unstable above the threshold, in the sense that the instability is a next-to-leading order effect. This complementary behavior of the instability is also reflected in the thermodynamic stability analysis where the specific heat capacity and isothermal permittivity show a similar behavior. It would be interesting to investigate the relation between the two approaches in more detail, that is, establish a more precise connection to the correlated stability conjecture \cite{Harmark:2005jk}. Also, it would be interesting for comparison to perform a numerical analysis of the long wavelength perturbations in the current setting as was done in the case of the neutral brane, where excellent agreement was found. 

Regarding the stability analysis, it is also worth noting that since the value obtained for the bulk viscosity leads to violation of the $\zeta/\eta$ bound, one might question the validity of the stability analysis for the case of a black brane charged under a top-form gauge field examined in ref. \cite{Emparan:2011hg}. Here the dispersion relations were written down using the assumption that the $\zeta/\eta$ bound proposed by \cite{Buchel:2007mf} is saturated. \vspace{5mm}

An interesting computation, that has not been investigated in the blackfold literature, is the computation of the entropy current \`{a} la \cite{Bhattacharyya:2008xc}. Computing the entropy current could provide a consistency check of the transport coefficients and the framework. We hope to address this question in the future.

Another natural future direction of this work is the generalization to black branes charged under higher form gauge fields possibly in the presence of a dilaton field. Of particular interest to string theory this would include the black D$p$-branes that carry charge under a Ramond-Ramond field for which the case of the D3-brane would be included. The case of the D3-brane would also be interesting in the context of the AdS/CFT correspondence. This could namely help elucidate possible relations between the AdS and flat space case i.e. the connection between the blackfold approach and the fluid dynamical regime of AdS/CFT. In this regard, it would also be interesting to understand, in a systematic manner, how the map of ref. \cite{Caldarelli:2012hy} extends when matter fields are included. This could also work as a method for obtaining the second order transport coefficients. However, taking the computation to second order from first principles (as in this work) is also of interest.

Finally, it would be interesting to include a Chern-Simon term in the theory. This was considered in AdS fluid/gravity in the papers \cite{Banerjee:2008th, Erdmenger:2008rm, Son:2009tf}. However, we note that black brane solutions analogous to the generalized Gibbon-Maeda solution with such a term in the action is to our knowledge not known in the literature.


\section*{Acknowledgments}

We thank Niels A. Obers for useful discussions, supervision, and review of the draft. We also wish to thank Nordita for hospitality during the program ``The Holographic Way: String Theory, Gauge Theory and Black Holes".


\newpage
\appendix

\section{Reduction} \label{sec::reduction}

In the first part of this appendix we will show how the equation of motions for the general case of a reduction of an Einstein-Maxwell theory on an Einstein manifold can be obtained. In the second part we will provide the example of applying the procedure for $d=2$ on the zeroth order solution.

\subsection{Reduction of Einstein-Maxwell theory on an Einstein manifold}

We consider Einstein-Maxwell theory on a $D$-dimensional space of the form
\begin{equation}
\text{d}s^2=g_{\mu\nu}\text{d}x^\mu\text{d}x^\nu=\text{d}s_{(b)}^2+e^{2\psi(x_b)}\text{d}s_{(E)}^2 \,.
\end{equation}
Here $\text{d}s_{(b)}^2$ denotes the metric of the base manifold $\mathcal M_{(b)}$, $x_{(b)}^i$ denotes the coordinates on $\mathcal M_{(b)}$, $\psi$ is a function on $\mathcal M_{(b)}$ 
and $\text{d}s_{(E)}^2$ is the metric of an Einstein manifold $\mathcal M_{(E)}$ with coordinates $x_{(E)}^A$. Since $\mathcal M_{(E)}$ is an Einstein manifold, we have
\begin{equation}
d_E \thinspace R^{(E)}=R_{E} \thinspace g^{(E)} \,,
\end{equation}
where $d_E$, $g^{(E)}$, $R^{(E)}$ and $R_E$ are respectively the dimension, the metric, the Ricci tensor and (constant) curvature scalar of $\mathcal M_{(E)}$. Moreover we consider a gauge field (minimally coupled to gravity) $A_\mu$
which only depends on $x_{(b)}^i$ and only takes values along the base manifold $\mathcal M_{(b)}$. Schematically 
\begin{equation}
A_\mu(x)=A_i(x_{b}) \,.
\end{equation}
The action $S$ of the system is given by 
\begin{equation}
S=S_g+S_{\text{EM}}, \quad S_g=\int d^{D}x\sqrt{|g|}R, \quad S_{\text{EM}}=\int d^{D}x\sqrt{|g|}\left[-\frac{1}{4}F_{\mu\nu}F^{\mu\nu}\right] \,,
\end{equation}
where $R$ denotes the Ricci scalar of the full metric $g_{\mu\nu}$. We can now perform a reduction and integrate out $\mathcal M_{(E)}$, one finds
\begin{equation}
\begin{split}
S_g &\sim \int_{\mathcal M_{b}}d^{d_b}x_{b}\sqrt{|g_{b}|}e^{d_E\psi(x_{b})}\left\{R_b+R_E e^{-2\psi(x)}-d_E(d_E-1)(\nabla \psi)^2\right\} \,, \\ 
S_{\text{EM}} &\sim -\int_{\mathcal M_{b}}d^{d_b}x_{b}\sqrt{|g_{b}|}e^{d_E\psi(x_{b})} F_{ij}F^{ij} \,.
\end{split}
\end{equation}
Having worked out the reduced action, it is easy to work out the equations of motion. As usual, the resulting system is EM theory on $\mathcal M_b$ coupled to a dynamical scalar field
and a current. The EOMs are 
\begin{equation}
\begin{split}
R_{ij}^{(b)}&=\frac{1}{4}\left(2F_{i}^{\ k}F_{jk}-\frac{1}{D-2}g_{ij}^{(b)}F_{mn}F^{mn}\right)+d_E\left(\nabla_a \psi\nabla_b \psi+\nabla_a\nabla_b \psi\right) \,, \\
\square \psi &+ d_E\left(\nabla \psi\right)^2-\frac{F_{mn}F^{mn}}{4(D-2)}=\frac{R_E e^{-2\psi}}{d_E} \,, \\
\nabla_i F^{ij}&=d_E F^{kl}\nabla_l \psi \,.
\end{split}
\end{equation}

\subsection{Reduction of the zeroth order solution}

In this section we demonstrate how the reduction works for the $0^{\text{th}}$ order solution with (fluid) dynamics in two spatial directions (in other words, an ordinary boost in the $(\sigma_1,\sigma_2)$ direction). Now the base space is composed of the three fluid brane directions (one time $\sigma^0$ and two spatial directions, $(\sigma^1, \sigma^2)$ along with the radial 
direction $r$). The metric has the form
\begin{equation}
\text{d}s^2=h^B\left[\left(\eta_{ab}+\left(1-\frac{f}{h^N}\right)u_a u_b\right)\text{d}\sigma^a\text{d}\sigma^b+\frac{\text{d}r^2}{f} + r^2\text{d}\Omega_{(n+1)}^2 + \sum_{i=3}^p\left(\text{d}x_\parallel^i\right)^2 \right] \,,
\end{equation}
with $a,b=0,1,2$ and where $x_\parallel^i$, $i=3,...,p$ are the $p-2$ static brane directions. We now integrate out the transverse sphere and the $p-2$ brane directions. The functions $\psi$ and $\phi$ are given by 
\begin{equation}
\phi(r)=\psi(r)+2\log r=B\log h(r) \,.
\end{equation}
It is now straightforward to compute $\kappa$, $j^\mu$ and $\mathcal F^\mu_{\ \nu}$. Here $x^\mu$ denotes coordinates of the four dimensional base space $x^\mu=(\sigma^0,\sigma^1,\sigma^2,r)$. One finds
\begin{equation}
\begin{split}
\kappa&=-\frac{Bn^2}{2}\left(\frac{r_0}{r}\right)^{2n}\frac{\gamma_0\left(1+\gamma_0\right)}{r^2h^N(r)} \,, \\
j^\mu\partial_\mu&=\frac{n^2}{2B}\left(\frac{r_0}{r}\right)^{2n}\frac{\sqrt{N\gamma_0(1+\gamma_0)}}{h^{N-1}(r)}\left(1+\frac{2}{B}+\frac{p}{n}h(r)+2\left(\frac{r_0}{r}\right)^{n}\gamma_0\right)u^a\partial_{\sigma_a} \,, \\
\mathcal F^\mu_{\ \nu}\partial_\mu \otimes \text{d}x^\nu&=\frac{N\kappa}{B}\left(u^au_b+\left(1-\frac{2}{N}\delta^a_b\right)\right)\partial_{\sigma_a} \otimes \text{d}\sigma^b-\frac{2\kappa}{B}\partial_r \otimes \text{d}r \,.
\end{split}
\end{equation}
It is now possible to show that, as expected, the reduced system obeys the EOMs with these effective sources. The above sources get derivative corrections in the perturbative expansion.

\section{Coefficients of the large \texorpdfstring{$r$}{\textit{r}} expansions}
\label{sec:sCoeff}

In this section, we list the first set of large $r$ expansion coefficients of the metric and gauge field given in section \ref{sec:firstorder}.

\subsection*{Scalar sector}

Below is listed the first set of coefficients for the large $r$ expansions of $f_{rv}$,
\begin{equation}
	\alpha_{rv}^{(1)} = - \frac{n((n+p)^2 + (n+p)(2(p+1)+n(p+2))\gamma_0 + 2p(p+2)\gamma_0^2)}{(n-1)(n+p)^2((n+1)+pB\gamma_0)} \,.
\end{equation}
Below is listed the first set of coefficients for the large $r$ expansions of $a_v$,
\begin{align}
\begin{split}
	\alpha_{v}^{(1)} &= \frac{n((2n+1)(n+p)^2 + (n+p)(1+n(2n+3)(p+1))\gamma_0 + 2p(1-p+2n(p+1)\gamma_0^2)}{(n-1)(2n-1)(n+p)^2((n+1)+pB\gamma_0)}\,, \\
	\beta_{v}^{(1)} &= \frac{(1+\gamma_0)^{\frac{N}{2}}}{n} \left[ 1 - p \gamma_0 B \left[ \frac{2(n+1)+pB\gamma_0}{((n+1)+ pB\gamma_0)^2} \right] \right] \,.
\end{split}
\end{align}
Below is listed the first set of coefficients for the large $r$ expansions of $f_{vv}$,
\begin{align}
\begin{split}
	\alpha_{vv}^{(1)} &= \frac{1}{(n-1)} \left( n(1 + 2\gamma_0) + \frac{4\gamma_0(n+p(1+\gamma_0)) }{(n+p)^2((n+1)+pB\gamma_0))} \right)  \,, \\
	\beta_{vv}^{(1)} &=  -(1+\gamma_0)^{\frac{N}{2}}  \frac{2(n+1)}{n(n+1 + pB \gamma_0)^2} \,.
\end{split}
\end{align}

\subsection*{Vector sector}

Below is listed the first set of coefficients for the large $r$ expansions of $f_{vi}$,
\begin{equation}
\begin{split}
	\alpha^{(1)}_{vi} &= (\partial_v \beta_i) \gamma_0 \left[- \frac{(n+p+1)(p+n(n+p+1)(1+2\gamma_0)}{(n-1)(n+p)^2} \right] + (\partial_i \gamma_0) \left[ - \frac{n+p+1}{(n-1)(n+p)} \right] \,, \\
	\alpha^{(2)}_{vi} &= \frac{n+p+1}{2(n-1)(2n-1)(n+p)^3} \bigg[ \\ 
	& \quad \bigg( 2(n+p)(n(n+p)+(4n^2+n-1-2p+4np)\gamma_0) \bigg) (\partial_i \gamma_0) \\
	& \quad \bigg( \gamma_0(4n(n+p)^2 + (n+p)(-1-2p+n(-3+2p+4n(4+3n+3p)))\gamma_0 \\
	& \quad + 4n(1+n+p) (-1+n+4n^2-2p+4np)\gamma_0^2 \bigg) (\partial_v \beta_i) \bigg] \,, \\
	\beta^{(1)}_{vi} &= 0  \,, \\
	\beta^{(2)}_{vi} &= - \frac{N}{4n} \left( \frac{2\gamma_0(1+\gamma_0) (\partial_v u_i) + (\partial_i \gamma_0)}{(1+\gamma_0)^{\frac{B}{2}} (1 + N \gamma_0) } \right) \,.
\end{split}
\end{equation}
Below is listed the first set of coefficients for the large $r$ expansions of $b_{i}$,
\begin{align}
\begin{split}
	\alpha^{(1)}_{i} &= \frac{1}{2(n-1)} \left[ \left( \frac{2(n+p+2n(n+p+1)\gamma_0)}{n+p} \right) (\partial_v \beta_i) + \left( \frac{1+2\gamma_0}{\gamma_0(1+\gamma_0)} \right) (\partial_i \gamma_0)  \right] \,. \\
	\beta^{(1)}_{i} &= \beta^{(2)}_{vi} \left[ \frac{n+p}{(n+p+1)\gamma_0(1+\gamma_0)}  \right]
\end{split}
\end{align}

\section{Thermodynamic coefficients} \label{sec:constR}

In this appendix we list a number of thermodynamic coefficients related to the analysis of section \ref{sec:stability}. The two coefficients $\mathcal{R}_1$ and $\mathcal{R}_2$ are given by
\begin{equation}
\begin{split}
	\mathcal{R}_1 &= \mathcal{Q}^2 \left[ \left(\frac{\partial \mathcal{Q}}{\partial \mathcal{T} }\right)_\Phi \left(\frac{\partial \epsilon}{\partial \Phi}\right)_\mathcal{T} - \left(\frac{\partial \mathcal{Q}}{\partial \Phi} \right)_\mathcal{T} \left(\frac{\partial \epsilon}{\partial \mathcal T}\right)_\Phi \right]^{-1} \,, \\
	\mathcal{R}_2 &= - \mathcal{R}_1 \left[ \mathcal{T} \left(\frac{\partial \epsilon}{\partial \mathcal{T} }\right)_\Phi + \Phi \left(\frac{\partial \epsilon}{\partial \Phi}\right)_{\mathcal{T}}  \right] \,.
\end{split}
\end{equation}
Writing out the speed of sound given in equation \eqref{eqn:speedofsound} it takes the form
\begin{equation} \label{eqn:csGeneral}
c_s^2= \frac{\mathcal{R}_1}{ \mathcal{Q}^2 w} \left[ w \left( \mathcal Q  \left(\frac{\partial \mathcal{Q}}{\partial \mathcal{T} }\right)_\Phi-s \left( \frac{\partial \mathcal{Q}}{\partial \Phi} \right)_\mathcal{T} \right) - \mathcal Q \left( \mathcal Q \left(\frac{\partial \epsilon}{\partial \mathcal T}\right)_\Phi -s \left(\frac{\partial \epsilon}{\partial \Phi}\right)_\mathcal{T}\right) \right] \,.
\end{equation}
Finally the coefficient associated to the dispersion relation of the sound mode is given by
\begin{equation}
\mathcal R= -\frac{1}{2} \frac{\mathcal{R}_1^2}{\mathcal{Q}^2 w^3 c_s^2} \left(\mathcal Q \left(\frac{\partial \epsilon}{\partial \mathcal{T}}\right)_\Phi - s \left(\frac{\partial \epsilon}{\partial \Phi}\right)_\mathcal{T} \right)
\left( \mathcal{Q} \frac{\mathcal{R}_2}{\mathcal{R}_1} +
w \left( 
\left( \frac{\partial \mathcal{Q}}{\partial \Phi} \right)_{\Phi} 
\Phi  +
\left( \frac{\partial \mathcal{Q}}{\partial \mathcal{T}} \right)_{\Phi} 
\mathcal{T}
\right)
\right) \,.
\end{equation}
For the Reissner-Nordstr\"om solution we have
\begin{equation}
\begin{split}
	\frac{\mathcal{R}_1}{\mathcal{T}} &= \frac{N\gamma_0}{n+1+pB\gamma_0},  \quad \frac{\mathcal{R}_2}{s \mathcal{T} \Phi} = \frac{1 - N \gamma_0 (1+2 \gamma_0) + n (1+N \gamma_0)^2}{1+2 \gamma_0 + n (1 - B \gamma_0)} \,, \\
	\mathcal{R} \frac{ w^2 }{s \mathcal{T}^2} &=  -\frac{2 N^2 \gamma_0^2(1+\gamma_0)^2  }{(1-B\gamma_0)(n+1+pB\gamma_0)} \,.
\end{split}
\end{equation}

\addcontentsline{toc}{section}{References}
\bibliographystyle{newutphys}
\bibliography{FluidGravity}

\end{document}